\documentclass[12pt]{article}
\usepackage{amssymb}
\usepackage{epsfig}
\usepackage{bbm}

\usepackage{graphicx}

\def\L{\mathcal L}
\def\e{\varepsilon}

\newcommand{\eqn}[1]{eq.~(\ref{#1})}

\newcommand{\wt}{\widetilde}
\newcommand{\mathsym}[1]{{}}

\newcommand{\p}{\varphi}

\begin{document}

\def\a{\alpha}
\def\b{\beta}
\def\c{\chi}
\def\d{\delta}
\def\e{\epsilon}
\def\f{\phi}
\def\g{\gamma}
\def\h{\eta}
\def\j{\psi}
\def\k{\kappa}
\def\l{\lambda}
\def\m{\mu}
\def\n{\nu}
\def\o{\omega}
\def\p{\pi}
\def\q{\theta}
\def\r{\rho}
\def\s{\sigma}
\def\t{\tau}
\def\u{\upsilon}
\def\x{\xi}
\def\z{\zeta}
\def\D{\Delta}
\def\F{\Phi}
\def\G{\Gamma}
\def\J{\Psi}
\def\L{\Lambda}
\def\O{\Omega}
\def\P{\Pi}
\def\Q{\Theta}
\def\S{\Sigma}
\def\U{\Upsilon}
\def\X{\Xi}
\def\Rmsma{r}
\def\Rmsmasq{r^2}
\def\ms{m_{\rm sol}}
\def\ma{m_{\rm atm}}
\def\ve{\varepsilon}
\def\vf{\varphi}
\def\vr{\varrho}
\def\vs{\varsigma}
\def\vq{\vartheta}

\def\nn{\nonumber}
\def\dg{\dagger}                                     
\def\ddg{\ddagger}                                   
\def\wt#1{\widetilde{#1}}                    
\def\mt{\widetilde{m}_1}
\def\mti{\widetilde{m}_i}
\def\rt{\widetilde{r}_1}
\def\mtt{\widetilde{m}_2}
\def\mttt{\widetilde{m}_3}
\def\rtt{\widetilde{r}_2}
\def\mb{\overline{m}}
\def\VEV#1{\left\langle #1\right\rangle}        
\def\be{\begin{equation}}
\def\ee{\end{equation}}
\def\ds{\displaystyle}
\def\ra{\rightarrow}
\def\dd{\displaystyle}
\def\bea{\begin{eqnarray}}
\def\eea{\end{eqnarray}}
\def\NO{\nonumber}
\def\Bar#1{\overline{#1}}

\def\lsim{\ \rlap{\raise 3pt \hbox{$<$}}{\lower 3pt \hbox{$\sim$}}\ }
\def\gsim{\ \rlap{\raise 3pt \hbox{$>$}}{\lower 3pt \hbox{$\sim$}}\ }
\def\cY{{\cal Y}}
\def\hcY{\hat{\cal Y}}


\def\pl#1#2#3{Phys.~Lett.~{\bf B {#1}} ({#2}) #3}
\def\np#1#2#3{Nucl.~Phys.~{\bf B {#1}} ({#2}) #3}
\def\prl#1#2#3{Phys.~Rev.~Lett.~{\bf #1} ({#2}) #3}
\def\pr#1#2#3{Phys.~Rev.~{\bf D {#1}} ({#2}) #3}
\def\zp#1#2#3{Z.~Phys.~{\bf C {#1}} ({#2}) #3}
\def\cqg#1#2#3{Class.~and Quantum Grav.~{\bf {#1}} ({#2}) #3}
\def\cmp#1#2#3{Commun.~Math.~Phys.~{\bf {#1}} ({#2}) #3}
\def\jmp#1#2#3{J.~Math.~Phys.~{\bf {#1}} ({#2}) #3}
\def\ap#1#2#3{Ann.~of Phys.~{\bf {#1}} ({#2}) #3}
\def\prep#1#2#3{Phys.~Rep.~{\bf {#1}C} ({#2}) #3}
\def\ptp#1#2#3{Progr.~Theor.~Phys.~{\bf {#1}} ({#2}) #3}
\def\ijmp#1#2#3{Int.~J.~Mod.~Phys.~{\bf A {#1}} ({#2}) #3}
\def\mpl#1#2#3{Mod.~Phys.~Lett.~{\bf A {#1}} ({#2}) #3}
\def\nc#1#2#3{Nuovo Cim.~{\bf {#1}} ({#2}) #3}
\def\ibid#1#2#3{{\it ibid.}~{\bf {#1}} ({#2}) #3}

\def\dd{\displaystyle}
\def\ud{\dd\frac{1}{\sqrt{2}}}
\def\ut{\dd\frac{1}{\sqrt{3}}}
\def\us{\dd\frac{1}{\sqrt{6}}}

\title{
%
\bf Flavor symmetries, leptogenesis  and the absolute neutrino mass scale}
\author{{\large E.~Bertuzzo$^a$, P.~Di Bari$^{b,c}$, F.~Feruglio$^{b}$, E.~Nardi$^{d}$}
\\
$^a$
{\it\small Scuola Normale Superiore and INFN,  Piazza dei cavalieri 7, 56126 Pisa, Italy}
\\
$^b$
{\it\small INFN, Sezione di Padova},
{\it\small Dipartimento di Fisica G.~Galilei}
{\it\small  Via Marzolo 8, 35131 Padua, Italy}
\\
$^c$
{\it\small School of Physics and Astronomy},{\it\small University of Southampton,}
{\it\small  Southampton, SO17 1BJ, U.K.}\\
{\it\small Department of Physics and Astronomy},{\it\small University of Sussex,}
{\it\small  Brighton, BN1 9QH, U.K.}
\\
 $^d$
  {\it\small INFN, Laboratori Nazionali di Frascati, C.P. 13, I00044     Frascati, Italy}
 \\
}

\maketitle \thispagestyle{empty}

\vspace{-7mm}

\begin{abstract}
  We study the interplay between flavor symmetries and leptogenesis in
  the case when the scale of flavor symmetry breaking is higher than
  the scale at which lepton number is violated.  We show that when the
  heavy Majorana neutrinos belong to an irreducible representation of
  the flavor group, all the leptogenesis $C\!P$ asymmetries vanish in
  the limit of exact symmetry.  In the case of reducible
  representations we identify a general condition that, if satisfied,
  guarantees the same result.  We then focus on the case of a model in
  which an $A_4$ flavor symmetry yields a drastic reduction in the
  number of free parameters, implying that at leading order several
  quantities are only a function of the lightest neutrino mass $m_l$,
  which in turn is strongly constrained.  For normal ordering (NO) we
  find $m_l\simeq (0.0044 \div 0.0056)\,{\rm eV}$ while for inverted
  ordering (IO) $m_l\gtrsim 0.017\,{\rm eV}$. For the $0\nu2\beta$
  decay parameter this yields $|m_{ee}|\simeq (0.006\div0.007)\,$eV
  (NO) and $|m_{ee}|\gsim 0.017\,$eV (IO).  We show that the
  leptogenesis $C\!P$ asymmetries only depend on $m_l$, on a single
  non-hierarchical Yukawa coupling $y$, and on two parameters that
  quantify the flavor symmetry breaking effects, and we argue that the
  unflavored regime for leptogenesis is strongly preferred in our
  model, thus realizing a rather predictive scenario.  Performing a
  calculation of the matter-antimatter asymmetry we find that for NO
  the observed value is easily reproduced for natural values of the
  symmetry breaking parameters. For IO successful leptogenesis is
  possible for a limited choice of the parameters implying
rather large reheating temperatures
$T_{\rm reh}\gtrsim 5\times 10^{13}\,{\rm GeV}$.

\end{abstract}

\newpage

\section{Introduction}

Perhaps the simplest explanation of the smallness of neutrino masses is in terms of the largeness of the scale at which
$B-L$ is violated, through the well-known seesaw mechanism \cite{seesaw} that, in the three neutrino framework, also implies $C\!P$ violation in the
lepton sector. Barring very special cancelations such a large scale is expected to be well above the weak scale, possibly near or just below the
grand unified scale. It is an intriguing coincidence that also baryogenesis, if occurring at temperatures well above the weak scale,
needs $B-L$ violation, since any initial $B+L$ asymmetry would be erased by sphaleron interactions in the subsequent
evolution of the universe.
The seesaw mechanism and baryogenesis elegantly combine in leptogenesis \cite{fy}
where the observed baryon asymmetry is produced
by the out-of equilibrium, $C\!P$ and $B-L$ violating decays of the right-handed (RH) neutrinos.
It is also quite remarkable that, at least in its simplest implementation, leptogenesis requires light neutrino masses below the eV scale \cite{nubounds,giudice,pedestrians},
in a range which is fully compatible with  other experimental constraints.
Unfortunately, it is difficult to promote this elegant
picture into a testable theory, due to the large number of independent parameters of any seesaw model.

For this reason it is of great interest to consider more constrained
frameworks, where, by restricting the number of relevant parameters,
the observed baryon asymmetry can be related to other observable
quantities.  Examples of such frameworks are models of lepton masses
where the total number of parameters is reduced by the presence of a
flavor symmetry.  There is a vast variety of models of this type: the
flavor symmetry can be global or local, continuous or discrete and
there are many viable candidate groups. A common feature of these
models is that the symmetry is always broken, to provide a realistic
description of the lepton masses and mixing angles.  The breaking can
be either explicit or spontaneous and is generally controlled by a set
of dimensionless symmetry breaking quantities $\eta$.  In general any
such model can be thought of as an expansion in powers of
$\eta$. There is a leading order approximation, corresponding to the
limit of exact flavor symmetry, plus small corrections proportional to
$\eta$ and its positive powers.  If $\eta$ are small, it makes sense
to neglect high powers of $\eta$ and the model predictions will
effectively depend on relatively few parameters.  Many models of
lepton masses based on flavor symmetries reproduce the most relevant
features of the lepton spectrum.  The main idea is to describe small
dimensionless observable quantities, such as charged lepton mass
ratios, $\theta_{13}$, $\theta_{23}-\pi/4$, in terms of some positive
power of the symmetry breaking parameters $\eta$.  In the context of
leptogenesis, given the extreme smallness of the baryon asymmetry, it
is quite natural to expect that the $C\!P$ asymmetries in the RH
neutrino decays are also suppressed by some power of $\eta$, opening
the interesting possibility of relating, through specific flavor
symmetries and flavor symmetry breaking patterns, the baryon asymmetry
to others low-energy observables.  Indeed if these $C\!P$ asymmetries
were of order $\eta^0$, we would need an abnormal suppression coming
either from a dilution factor from photon production or from the
wash-out factor. Typically we expect a dilution factor of order
$10^{-2}$ and a wash-out factor in the range $10^{-3}\div 10^{-2}$,
which favor $C\!P$ asymmetries around $10^{-6}\div 10^{-5}$.

We think that establishing the dependence of the $C\!P$ asymmetries on $\eta$ could be of great interest for model building and for
phenomenology. In this paper we collect a number of observations that can be useful to this scope.
In the first part of this work, section 2, we consider a completely general framework
and we derive conditions for the vanishing of the total $C\!P$ asymmetries based on
general group theoretical arguments.
The main result of this part is the proof that when the heavy Majorana neutrinos
are assigned  to an irreducible representation of the flavor group, all the
leptogenesis $C\!P$ asymmetries vanish in the limit of exact symmetry.
A general condition for the vanishing of the leptogenesis $C\!P$ asymmetries
that hold for different types of flavor symmetries is also given.

In the second part of the paper we work out in detail the baryon asymmetry
predicted by a model of lepton masses based on the $A_4$ discrete symmetry group ~\cite{A4Z3},
originally built to reproduce the tri-bimaximal lepton mixing (TBM),
a good approximation of the existing data \cite{Harrison:2002er}. In such a model
there are strong indications that leptogenesis can be successfully realized,
since the leptogenesis $C\!P$ asymmetries are of order $\eta^2$ \cite{Jenkins:2008rb}
and the $\eta$ parameters are of order $10^{-2}$.
However, an estimate of the washout effects is still lacking and, given the small number of
independent parameters, it is not a priori guaranteed that the baryon asymmetry can be
reproduced.

In section 3 we review the main features of the $A_4$ model.  We show how this
model is surprisingly predictive: if the neutrino ordering is normal,
the neutrino mass spectrum and the $0\nu 2\beta$ decay parameter
$|m_{ee}|$ are determined rather precisely.  If the ordering is
inverted, interesting lower limits on both quantities are obtained.
In section 4 we confront the $A_4$ model with leptogenesis.
We first compute the leptogenesis $C\!P$ asymmetries at sub-leading order.
Next, both for the normal and inverted ordering cases, we estimate the
value of the cosmic baryon asymmetry resulting from leptogenesis.
It is quite surprising that in the case of normal hierarchy the right
values for the $C\!P$ asymmetries are obtained despite the
tight constraints found on the involved parameters, namely on the
phases and on the absolute neutrino mass scale. In the case of inverted
ordering successful leptogenesis is possible but it relies on a compensation
between very large $C\!P$ asymmetries $|\ve_i|\sim 10^{-2}$ and strong
wash-out from $\Delta L=2$ processes. This compensation occurs for
very large RH neutrino masses $\sim 10^{14}\,{\rm GeV}$ implying in turn
values of the reheating temperature not lower than $\sim 5\times 10^{13}\,{\rm GeV}$
for quasi-degenerate light neutrinos, only marginally compatible with cosmological observations.
Finally, in section 5 we collect our main results and
draw the conclusions.

\section{Flavor symmetries and leptogenesis $C\!P$ asymmetries}

In this section we discuss the constraints on the leptogenesis $C\!P$
asymmetries implied by different classes of flavor symmetries.  We
assume that the flavor symmetry is broken at a scale above the
leptogenesis scale, that is, leptogenesis occurs in the standard
framework of the SM plus three heavy Majorana seesaw
neutrinos~\cite{seesaw}, without involving in its dynamics any
additional state.\footnote{In the cases when the flavor symmetry is
  broken close or below the leptogenesis scale, interesting scenarios
  that are quite different from the standard one, and that in general
  involve the dynamics of new scalar or fermion particles, might be
  realized~\cite{AristizabalSierra:2007ur}.}  The most interesting
situations occur when the leptogenesis $C\!P$ asymmetries vanish in the
limit of exact flavor symmetry.  Since in several models the breaking
of the symmetry is related in a clear way to parameters that are
measurable at low energy (like e.g. $\theta_{13}$ or
$\pi/4-\theta_{23}$) our results open up the possibility of
constraining - within specific flavor models - the leptogenesis $C\!P$
asymmetries (that are the crucial high energy parameters) from low
energy experiments. For definiteness we refer to the non-supersymmetric
case but all our results hold in the supersymmetric case as well which will
be considered in Section 4 within a particular relevant example.

\subsection{The leptogenesis $C\!P$ asymmetries}

We write the Lagrangian for the three left-handed lepton doublets $L_{\alpha}$,
the three RH charged leptons $E_\alpha$, and  the three
heavy singlet Majorana neutrinos $N_i$ as:
\begin{equation}
\label{eq:L}
-{\cal L} = \frac{1}{2} N_i M_{ij} N_j +
\bar L_\alpha Y_{\alpha i} N_i H_u+\bar L_{\alpha} Y^e_{\alpha\beta} E_\beta H_d
\end{equation}
where $H_{u,d}$ are the $SU(2)$ doublets Higgs fields, the matrix
$M_{ij}$ is complex symmetric, and the Yukawa matrices $Y_{\alpha i}$ and
$Y^e_{\alpha\beta}$ are arbitrary complex $3\times 3$ matrices.

The leptogenesis $C\!P$ asymmetries for $N_i$-decays
into $\alpha$-leptons ($\alpha=e,\mu,\tau$) are defined as
\be
\ve_{i\ra\a}\equiv  \, {\G_{i\ra\alpha}-\overline{\G}_{i\ra\alpha}
\over \G_{i}+\overline{\G}_{i}} \, ,
\ee
where $\Gamma_i\ra \a$ and $\bar{\Gamma}_i\ra \a$ denote respectively the
decay rates into $\a$-leptons and $\a$-anti-leptons. We have defined
$\Gamma_i\equiv \sum_\alpha \G_{i\ra\alpha}$ (with an analogous definition for
$\overline{\G}_{i}$) so that the total $N_i$-decay widths are given by
$\Gamma_i+\bar{\Gamma}_i$, and the total  $C\!P$ asymmetries $\varepsilon_i\equiv
\sum_\alpha \varepsilon_{i\to \alpha}$ are
\be
\ve_i\equiv  \,{\G_i-\bar{\G}_i\over \G_i+\bar{\G}_i} \,.
\ee

Let us now introduce the Hermitian matrix
\begin{equation}
  \label{eq:YY}
  \cY_{ij} \equiv \left(Y^\dagger Y\right)_{ij}, \qquad   \cY =  \cY^\dagger.
\end{equation}
A one-loop perturbative calculation of the flavor and of the total CP
asymmetries $\varepsilon_{i\to \alpha}$ and $\varepsilon_i$
gives~\cite{co96}:
\begin{eqnarray}
  \label{eq:flCPasymm}
 \varepsilon_{i \to \alpha} &=&
\frac{1}{8\pi \hcY_{ii}}\sum_{j\neq i}
\left\{ {\rm Im}\left[\hat Y^*_{\alpha i} \hcY_{ij} \hat Y_{\alpha j}\right] f_{ij} +
{\rm Im}\left[\hat Y^*_{\alpha i} \hcY_{ji} \hat Y_{\alpha j}\right] g_{ij}\right\} \\
  \label{eq:CPasymm}
 \varepsilon_{i} &=& \sum_\alpha
 \varepsilon_{i \to \alpha} =
\frac{1}{8\pi \hcY_{ii}}\sum_{j\neq i}
 {\rm Im} \left[\hcY_{ij}^2\right] f_{ij} \, ,
  \end{eqnarray}
  where the hat denotes the basis in which the matrices $M$ and $Y^e$ of
  eq.~(\ref{eq:L}) are diagonal, with real and non-negative entries.  The
  $f_{ij},\, g_{ij}$'s are known \cite{co96} functions of the heavy Majorana
  neutrino masses $M_i$.  Indicating with $M_l$ the lightest heavy neutrino mass
  ($M_l=\min\{M_1,M_2,M_3\}$), defining $x_i\equiv M^2_i/M_l^2$ and introducing
  the function
\be
\label{xi} \xi(x)= {2\over 3}\,x\,
  \left[{2-x\over 1-x}-
(1+x)\,\ln\left({1+x\over x}\right)
\right] \, ,
\ee
  such that $\xi(\infty)=-1$, they are given by
\be
\label{eq:fgij}
  f_{ij}={3\over 2}\,\frac{\x(x_j/x_i)}{\sqrt{x_j/x_i}} \hspace{10mm}
  \mbox{and} \hspace{10mm} g_{ij}=\frac{1}{1-x_j/x_i} \,.
\ee
  In the limit of strongly hierarchical singlet neutrinos $x_j\gg
  x_i$, one has $|\,g_{ij}|\ll \vert f_{ij}\vert \sim
  (3/2)\sqrt{x_i/x_j})$ and, barring strong phase cancelations in the
  Yukawa couplings, the second term in $\varepsilon_{i \to \alpha}$
  can be neglected.  In the quasi-degenerate limit $x_j/x_i\to 1$ we
  have $|g_{ij}| \approx |f_{ij}| \sim |1-x_j/x_i|^{-1}$.  It is
  apparent from eq.~(\ref{eq:flCPasymm}) that if $\hcY$ is a diagonal
  matrix, all the flavor dependent CP asymmetries vanish.  As regards
  the total CP asymmetries in eq.~(\ref{eq:CPasymm}), they vanish also
  in the case $\hcY$ has non-vanishing but real off-diagonal entries.

\subsection{Flavor symmetries}

We assume that the theory is invariant under transformations of a flavor
symmetry group $G$, spontaneously broken down to a subgroup $H$, through the
vacuum expectation values (VEVs) of a set of scalar fields $\varphi$. The
ratios between $\langle \varphi\rangle$ and the ultraviolet cutoff $\Lambda$
of the theory provide a set of small parameters $\eta\simeq \langle
\varphi\rangle/\Lambda$, in terms of which we can expand the quantities of
interest. The symmetry group $G$ can be discrete or continuous, global or
local. The group $G$  acts on the 3 left-handed electroweak doublets $L$, 3
RH neutrinos $N$ and 3 RH charged leptons $E$,
as follows
\begin{equation}
L'=\Omega_L(g) L~~~,~~~~~~~N'=\Omega_N(g) N~~~,~~~~~~~E'=\Omega_E(g) E~~~~,
\end{equation}
where $\Omega_{I}(g)$ $(I=L,N,E)$ denote unitary representations of
the group $G$ for the generic group element $g$. After breaking of the
group $G$, the Lagrangian \eqn{eq:L} is generated, and the three
matrices $M_{ij}$, $Y_{\alpha i}$ and $Y^e_{\alpha\beta}$ will be in
general functions of the symmetry breaking parameters $\eta$.  To
study the implications of flavor symmetries for the leptogenesis $C\!P$
asymmetries it is then convenient to expand these matrices in powers
of $\eta$:
\begin{equation}
\label{eq:expansion}
M=M^0+\delta M+\dots~,
~~~~~~Y=Y^0+\delta Y+\dots~,
~~~~~~Y^e=Y^{e0}+\delta Y^e+\dots~
\end{equation}
In these expansions $M^0$, $Y^0$ and $Y^{e0}$ denote the terms
surviving in the limit of exact symmetry $\eta\to 0$, while $\delta
M$, $\delta Y$ and $\delta Y^e$ are all terms ${\cal O}(\eta)$.
Flavor symmetries will imply different constraints on these matrices as
well as on the matrix $\cY$ of~\eqn{eq:YY}, depending on their nature
(non-Abelian or Abelian) and depending also on the specific
representations to which the fields are assigned. We now analyze the
different possibilities.

\smallskip
\subsubsection {Irreducible representations $\Omega_N$}
\label{sec:irreducible}
We start by considering the case where $\Omega_N$ is a
three-dimensional irreducible representation of a non-Abelian group
$G$. In this case the implications of the flavor symmetry for
leptogenesis are simple and completely general: that is, in the limit
of exact symmetry $\eta\to 0$ all the leptogenesis $C\!P$ asymmetries
vanish.

Let us consider the matrix $\cY$ of eq.~(\ref{eq:YY}) in the limit
$\eta\to 0$.  Due to the invariance under $G$, $\cY^0\equiv
{Y^0}^\dagger Y^0$ obeys the relation:
\begin{equation}
\Omega_N^\dagger (g)\, \cY^0\, \Omega_N(g)=\cY^0
\end{equation}
for any group element $g$. Therefore $\cY^0$ commutes with $\Omega_N(g)$ for
any $g$. Then, by the first Shur's lemma, $\cY^0$ is a multiple of the
identity matrix $\cY^0 = |y|^2\,I $, and $Y_0$ itself is proportional to
a unitary matrix.  This property is clearly basis-independent and holds in
particular in the hatted basis, leading to the vanishing of all the $C\!P$
asymmetries in the limit of exact symmetry:
\begin{equation}
 \varepsilon_{i \to\alpha}= \varepsilon_{i}= 0\qquad {\rm for} \qquad \eta =
  0.
\label{CPirr}
\end{equation}
Clearly, to accommodate the three $N$'s in an irreducible
representation, $G$ should be non-Abelian, since Abelian groups have
only one-dimensional irreducible representations.  To study the
dependence of the $C\!P$ asymmetries on $\eta$ when the flavor symmetry is
not exact, we expand the matrix $\cY$ in the hatted basis obtaining:
\begin{equation}
\hcY_{ij} = |y|^2 \delta_{ij} + \delta \hcY_{ij}, \qquad \qquad
\delta \hcY=
{\hat {Y^0}}^\dagger { \delta\hat Y}+{\delta\hat Y}^\dagger {\hat Y^0} +\dots
\end{equation}
Inserting this in the eq.~(\ref{eq:flCPasymm}), and assuming for the time
being that at the leading order $f_{ji}$ and $g_{ji}$ in eq.~(\ref{eq:fgij})
are unrelated to $\eta$, we obtain:
\begin{equation}
\label{eq:Oeta}
\varepsilon_{i \to \alpha} \sim \sum_{j\neq i}  {\rm Im} \left[\hat Y^*_{\alpha i}
\hat Y_{\alpha j}\,  \delta \hcY_{ij}\right] \sim {\cal O}(\eta).
\end{equation}
This is because while $ \delta \hcY$ is ${\cal O}(\eta)$, $Y^*_{\alpha
  i}Y_{\alpha j}$ is not related to $\eta$ when no sum over $\alpha$ is taken.
In contrast, by summing over $\alpha$ we obtain
%
\begin{equation}
\label{eq:Oeta2}
\varepsilon_{i} \sim \sum_{j\neq i}  {\rm Im} \left[
\delta \hcY_{ij}^2
\right] \sim {\cal O}(\eta^2).
\end{equation}
Thus we can conclude that when the heavy Majorana neutrinos $N_i$ belong to an
irreducible representation of a non-Abelian flavor symmetry, the following two
possibilities can be realized:
\begin{itemize}
\item[1.]  If $M_i\lsim 10^{12}$ GeV \cite{bcst,flavor1,flavor2,inverse,Davidson:2008bu}
\footnote{Notice that this condition becomes, more generally, $M_i\lsim 10^{12}\,(1+\tan^2\b)$ GeV in the
supersymmetric case \cite{antusch}.}, then
the role of lepton flavor dynamics cannot be neglected in
a description of leptogenesis from $N_i$-decays~
and the final asymmetry is a linear combination of the flavored $C\!P$ asymmetries
and therefore of first order in the symmetry breaking parameter $\eta$.
\item[2.] If $M_i\gsim 10^{12}$ GeV (see footnote),
  then the one flavor approximation is accurate, the relevant quantities are
  the total $C\!P$ asymmetries $\varepsilon_i$ and therefore the final
  asymmetry gets suppressed by one additional power of
  $\eta$ with respect to the previous case.
\end{itemize}
In the cases when the non Abelian symmetry also implies degeneracies
for the singlet neutrinos mass eigenvalues ($|M_i-M_j|\ll M_i+M_j$),
then we have $g_{ij}\sim f_{ij}\sim {\cal O}(\eta^{-1})$ and
eqs.~(\ref{eq:Oeta}) and (\ref{eq:Oeta2}) become $\varepsilon_{i\to
  \alpha} \sim {\cal O}(\eta^{0})$ and $ \varepsilon_{i} \sim {\cal
  O}(\eta)$.

The previous analysis applies to several popular models that predict TBM,
based for example on the discrete symmetry groups $A_4$~\cite{Ma:2004zv,Ma:2005mw}and
$Z_7\rtimes Z_3$~\cite{Luhn:2007sy} (that were analyzed
in~\cite{Jenkins:2008rb} in the context of leptogenesis), $T'$~\cite{Tprime},
$S_4$~\cite{S4}, $D(4)$ \cite{D4}, on the continuous group $O(3)$ that occurs
e.g. in models of minimal flavor violation~\cite{Cirigliano:2006nu}, etc.


\subsubsection {Reducible representations $\Omega_N$}
In the cases when the heavy Majorana neutrinos $N_i$ transform according to a
reducible representation $\Omega_N$ of the group $G$, it is still possible to
derive some general conclusions.  When at most one eigenvalue of $M^0$ vanishes,
$\hcY^0$ can be brought into diagonal form if
\begin{equation}
{\cY^0}^T M^0-M^0 {\cY^0}=0.
\label{com1}
\end{equation}
Clearly, this implies that also in this case the leptogenesis $C\!P$ asymmetries
vanish in the limit of exact symmetry.  Indeed, if $\hcY^0$ is diagonal in the
basis where $\hat M^0$ is diagonal, then there exists a unitary matrix $U$
that diagonalizes both matrices: $\hat M^0= U^*M^0U^\dagger $ and $\hcY^0= U \cY^0
U^\dagger$.  Then, from $[\hat M^0, \hcY^0]=0$, the condition (\ref{com1})
immediately follows.  Conversely, let us assume that eq. (\ref{com1}) is
satisfied.  Since $M^0$ (like $M$) is a symmetric matrix, there exist (Takagi
factorization) a unitary matrix $U$ such that
\begin{equation}
\hat{M}^0=U^*\,M^0\,U^\dagger~,
\label{takagi}
\end{equation}
where $\hat{M}^0$ is diagonal with real and non-negative eigenvalues.  By
acting with $U^T$ and $U$ respectively on the left and right-hand sides of
eq.~(\ref{com1}) (and recalling the definition $\hcY^0 = U \cY^0 U^\dagger$)
we obtain
\begin{equation}
(\hat{\cal Y}^0)^T \hat{M}^0-\hat{M}^0 \hat{\cal Y}^0
\label{com2}
\end{equation}
Since $\cY^T=\cY^*$, eq.~(\ref{com2}) implies for the $(ij)$ component of the
matrix product above
\begin{equation}
  \label{eq:components1}
{\rm Re}(\hcY^0_{ij}) \left[ M^0_{ii}-M^0_{jj}\right] = 0,
\qquad
\qquad
 {\rm Im}(\hcY^0_{ij}) \left[ M^0_{ii}+M^0_{jj}\right] = 0.
\end{equation}
From eq.~(\ref{eq:components1}) we can conclude the following:
\begin{itemize}
\item[1.]  If the eigenvalues of $M^0$ are all different
then $\hcY^0$ is diagonal. For the leptogenesis
  $C\!P$ asymmetries, the same conclusions stated at the end of the previous
  section hold.

\item[2.]  If two eigenvalues of $M^0$ are equal and nonvanishing, then
  $\hcY^0$ can  always be rotated into diagonal form.  To prove this, let us
  assume e.g. that $M^0_i=M^0_j\neq 0$. Then $\hcY^0$ is block diagonal with real
  entries in the corresponding $2\times 2$ block
  $\hcY^0_{(ij)}$.\footnote{Note that while this would imply that the total $C\!P$
    asymmetry $\varepsilon_{i}$ vanishes, it would not necessarily imply the
    same for the flavor asymmetries $\varepsilon_{i\to \alpha}$, see
    eq.~(\ref{eq:Oeta}).}  It is then possible to diagonalize $\hcY^0_{(ij)}$
  by means of a $2\times 2$ orthogonal transformation
  $O_{(ij)}O_{(ij)}^T=I_{2\times 2}$ while leaving $M^0_{(ij)}= M_i\,
  I_{2\times 2}$ unaffected.

\item[3.] If all the three eigenvalues of $M^0$ are equal and non-vanishing,
  then, following the argument outlined above, $\hcY^0$ can be always brought
  into diagonal form.

\end{itemize}
Since in cases 2. and 3.  the eigenvalues of $M$ are characterized by
degeneracies when the limit $\eta\to 0$ is taken, the conclusion that the off
diagonal terms of $\hcY_{ij}$ are at most of ${\cal O}(\eta)$ by itself is not
sufficient to infer an estimate of the size of the $C\!P$ asymmetries.  As was
already mentioned above, when $M_i = M_j+{\cal O}(\eta)$ we have that
$f_{ij},\, g_{ij} \sim {\cal O}(\eta^{-1})$.  Thus, only the total $C\!P$
asymmetry $\varepsilon_{i}$ gets parametrically suppressed by one power of
$\eta$, while for the flavored $C\!P$ asymmetries there is no parametric
suppression.

The last possibility, that is not contemplated in 1.--3., is that two
eigenvalues of $\hat{M}^0$ (or all the three) vanish, implying that two
(three) light neutrinos are massless in the symmetric limit.  In this case we
have no information on the corresponding block $\hcY^0_{(ij)}$ (on $\hcY^0$),
and to reach some conclusion, we need to analyze the structure of the
correction $\delta M$.  For example, if $M^0$ itself vanishes, and
\begin{equation}
{\cY^0}^T\, \delta M-\delta M\, {\cY^0}=0,
\label{com3}
\end{equation}
then again we can conclude that the off-diagonal terms $\hcY^0_{i\neq j}$ are
at least of ${\cal O}(\eta)$.
It is then clear that, in several relevant cases, in order to derive general
conclusions about the size of the leptogenesis $C\!P$ asymmetries the criterion
eq.~(\ref{com1}), can still be  applied  to terms of higher
order in $\eta$ in the expansions of  $M$ and $\cY$.

Popular models for which the conclusions of this section can be relevant
are based on the discrete symmetries $S_3$~\cite{S3}, $D_4$~\cite{D4},
$\mu-\tau$ permutation symmetry.

\subsubsection{The Abelian case}

A particular case  of a reducible representation $\Omega_N(g)$,
is when it corresponds to the  sum of three singlet representations:
\begin{equation}
\Omega_N(g)=
\left(
\begin{array}{ccc}
\omega_1(g)&0&0\\
0&\omega_2(g)&0\\
0&0&\omega_3(g)
\end{array}
\right)
\end{equation}
where $|\omega_i(g)|^2=1$. If this occurs, in practice,
the action of $G$ on $N$ is Abelian.

We can examine all possible cases. For simplicity, we assume that the
transformation properties under $G$ of the left-handed doublets $L_\alpha$ are
such that the maximum allowed number of entries in $Y$ are different from zero.
\begin{enumerate}
\item $\omega_i(g)=1~~~(i=1,2,3)$.\\
  All the elements of $M^0$ are non-vanishing and are free-parameters of the
  model, and so are the eigenvalues matrix $\hat M^0$.  In general, also the
  elements $\hcY_{i\neq j}$ will be non-zero, since the matrix elements of
  $M^0$ and those of $\cY$ are completely unrelated.  Concerning the $C\!P$
  asymmetries, the only conclusions that can can be inferred is that
   they will be generically non-vanishing.  Models based on Abelian symmetries
  of the Froggatt-Nielsen type often realize this situation.

\item  $\omega_2(g)=\omega_3(g)=1$ and $\omega_1(g)\ne 1$.\\
  $M^0$ is block diagonal with non-vanishing free parameters filling up
  $M^0_{(2,3)}$, and condition eq.~(\ref{com1}) in general is {\it not}
  realized.  If $\omega_1=-1$, then $M^0_{11}\neq 0$, otherwise $M^0_{11}$
  vanishes.  The invariance condition $\Omega_N^\dagger \cY^0 \Omega_N= \cY^0$
  implies $\hcY^0_{12}=\hcY^0_{13}=0$, and therefore all the $C\!P$ asymmetries
  for $N_1$ vanish. The $N_{2,3}$ $C\!P$ asymmetries are generically different
  from zero.

\item $\omega_3(g)=1$ and $\omega_1(g),\,\omega_2(g)\ne 1$.\\
  $M^0$ is block diagonal with $M^0_{33}\neq 0$.  If $\omega_1=\omega_2=-1$
  the $M^0_{(1,2)}$ block is non-vanishing with free parameters filling up all
  the entries. Clearly (modulo the replacement $N_{1} \leftrightarrow N_{3}$)
  this case reduces to the previous one, and in particular $\Omega_N^\dagger
  \cY^0 \Omega_N= \cY^0$ implies
  $\hcY^0_{31}=\hcY^0_{32}=0$ and the vanishing of the $N_3$ $C\!P$ asymmetries.

  If $\omega_1,\,\omega_2\neq \pm 1$ but $\omega_1\omega_2=1$, then the
  $M^0_{(1,2)}$ block is anti-diagonal. These two conditions also imply
  $\omega_1^*\omega_2\neq 1$ and thus in the flavor basis $\cY$ is a diagonal matrix.
  However, since condition (\ref{com1}) in general is not realized,
  in the hatted basis $\hcY_{12}\neq 0$  and the $C\!P$ asymmetries for $N_{1,2}$
  do not vanish.

\item If $\omega_i\neq \pm 1$ (for $i=1,2,3$) and $\omega_i\omega_j\neq 1$ for
  all $i\neq j$, then $M^0=0$ and the heavy neutrino masses arise at ${\cal
    O}(\delta M)$ or at higher order. Unless $\omega_i^*\omega_j= 1$ for some
  $i\neq j$, $\cY$ is diagonal in the flavor basis, but condition (\ref{com2})
  is generally not satisfied. Thus in the hatted basis $\hcY^0_{i\neq j}\neq
  0$ and all the $C\!P$ asymmetries are generically non-vanishing.

\end{enumerate}

\section{The flavor symmetry $A_4$}

In recent years, the non-Abelian discrete group $A_4$~\cite{Ma:2004zv}
stemmed out as one of the most interesting candidates for explaining
the measured values of neutrino mass square differences and mixing
angles within a TBM~\cite{Harrison:2002er} phenomenological pattern.
In models based on $A_4$ the heavy Majorana neutrinos are generally
assigned to an irreducible representation of the group.  Thus the main
conclusion of Sect.~\ref{sec:irreducible}, that is that
$\varepsilon_i,\, \varepsilon_{i\to \alpha} $ vanish in the limit of exact symmetry,
applies \cite{Jenkins:2008rb}. For this reason $A_4$ represents a relevant case in which the
constraints on the leptogenesis $C\!P$ asymmetries implied by the flavor
symmetry are highly non-trivial.  As we will discuss in this section,
besides this general property, the specific realization of $A_4$ on
which we will focus, and that is based on the flavor symmetry $A_4\times
Z_3\times U(1)_{FN}$, provides a highly constrained framework that results in some
remarkably precise predictions for the neutrino sector.

\subsection{The discrete group $A_4$}

We start by  reviewing briefly  the properties of the $A_4$ group.
$A_4$ is the alternating group of order $4$, e.g. the group of the
even permutations of four objects, so it has $12$ elements.
From a geometrical point of view, it is the subgroup of the
three-dimensional rotation group leaving invariant a regular tetrahedron.
All the elements of the group can be expressed in terms of only two
elements of the group itself (the generators) which we will call $S$ and $T$.
They obey the following rules:
\begin{equation}
 S^2=T^3=(ST)^3=1
\end{equation}
$A_4$ has four inequivalent irreducible representations (irreps from now on): three of dimension 1 ($1$, $1'$ and $1''$) and one of dimension 3 ($3$).\\
The form of the generators in the different irreps is given by:
\begin{eqnarray}
 1&:& S=1,~~~~~~~ T=1\\
1'&:& S=1, ~~~~~~~T=\omega^2\\
1''&:& S=1, ~~~~~~~T=\omega
\end{eqnarray}
\begin{equation}
 3:S=\frac 13\left(
\begin{array}{ccc}
 -1&2&2\\
2&-1&2\\
2&2&-1\\
\end{array}
\right),~~~~~~~
T=\left(
\begin{array}{ccc}
 1&0&0\\
0&\omega^2&0\\
0&0&\omega\\
\end{array}
\right)
\end{equation}\\
The product of two triplets decomposes as follows $$3 \times 3= 1+1'+1''+3_S+3_A$$
Explicitly, if we have two triplets $a=(a_1,a_2,a_3)$ and $b=(b_1,b_2,b_3)$, the product reads
\begin{equation}
 (ab)_{k}=\sum_{i,j}a_i A^k_{ij} b_j
\end{equation}
where $k=1, 1', 1'', 3_S, 3_A$. The $A^k$ matrices are given by:
\begin{equation}
\label{eq:uno}
 A^1=\left(
\begin{array}{ccc}
 1&0&0\\
0&0&1\\
0&1&0\\
\end{array}
\right),\qquad
%
A^{1'}=\left(
\begin{array}{ccc}
 0&1&0\\
1&0&0\\
0&0&1\\
\end{array}
\right),\qquad
%
A^{1''}=\left(
\begin{array}{ccc}
 0&0&1\\
0&1&0\\
1&0&0\\
\end{array}
\right)
\end{equation}

\begin{equation}\label{eq:tre_S}
A^{3_S}_1=\frac 13\left(
\begin{array}{ccc}
 2&0&0\\
0&0&-1\\
0&-1&0\\
\end{array}
\right), \
A^{3_S}_2=\frac 13\left(
\begin{array}{ccc}
 0&-1&0\\
-1&0&0\\
0&0&2\\
\end{array}
\right), \
A^{3_S}_3=\frac 13\left(
\begin{array}{ccc}
 0&0&-1\\
0&2&0\\
-1&0&0\\
\end{array}
\right)
\end{equation}

\begin{equation}\label{eq:tre_A}
A^{3_A}_1=\frac 12\left(
\begin{array}{ccc}
 0&0&0\\
0&0&1\\
0&-1&0\\
\end{array}
\right),\
A^{3_A}_2=\frac 12\left(
\begin{array}{ccc}
 0&1&0\\
-1&0&0\\
0&0&0\\
\end{array}
\right),\
A^{3_A}_3=\frac 12\left(
\begin{array}{ccc}
 0&0&-1\\
0&0&0\\
1&0&0\\
\end{array}
\right)
\end{equation}
where in eqs.~(\ref{eq:tre_S}) and~(\ref{eq:tre_A}) we have used the
convention that the subscript labels the component of the triplet, \emph{e.g.}
\begin{equation}
  \label{eq:components}
  (ab)_{3_A}=\frac 12 (a A_1^{3_A} b, a A_2^{3_A} b, a A_3^{3_A} b)=
\frac 12 (a_2 b_3 - a_3 b_2, a_1 b_2 -a_2 b_1, a_3 b_1 - a_1 b_3)
\end{equation}
The group $A_4$ has two obvious subgroups: $G_S \simeq Z_2$,
the reflection subgroup generated by $S$, and $G_T\simeq Z_3$,
which is generated by $T$.
It is immediate to see that the VEVs
$$\langle \varphi_T \rangle\propto (1,0,0)$$
$$\langle \varphi_S \rangle \propto (1,1,1)$$
break $A_4$ respectively to $G_T$ and $G_S$.
As we will see in the next section,
both triplets $\varphi_T$ and $\varphi_S$
are needed in order to get the TBM
matrix for the neutrinos.

\subsection{The $A_4\times Z_3\times U(1)_{FN}$ model}

We recall here the main features  of a well known model ~\cite{A4Z3}  for TBM
that is based on the symmetry group $A_4\times Z_3\times U(1)_{FN}$
\footnote{Several models based on the $A_4$ symmetry have been constructed.
Recently the baryon asymmetry was computed in one of these variants,
where the RH neutrino masses are strongly degenerate \cite{branco}.}.
In order to write the most general Lagrangian invariant under the symmetry
group, we first have to assign the fields to specific $A_4$ irreps.  We assume
that the model is supersymmetric, this turns out to be useful to ensure the
correct vacuum alignment of the scalar fields breaking $A_4$, that is a
crucial  feature to obtain TBM.
We divide the  field content of the model in three different sectors:
\begin{enumerate}
\item \emph{matter fields sector}: the lepton doublets $L_\alpha$ transform as
  a triplet $3$, while the RH charged leptons $E_1\equiv e^c$, $E_2\equiv
  \mu^c$ and $E_3\equiv \tau^c$ transform respectively as $1$, $1''$ and $1'$.
  In order to discuss leptogenesis, we need to introduce also the three heavy
  Majorana $SU(2)$ singlet neutrinos $N_i$, which are assigned to another
  triplet~$3$.
\item \emph{symmetry breaking sector}: the Higgs fields $H_u$ and $H_d$ (with
  opposite hypercharge) transform in the $1$ irrep of $A_4$.  We introduce
  four flavons for the spontaneous breaking of $A_4$ $\varphi_T\simeq 3$,
  $\varphi_S\simeq 3$, $\xi\simeq 1$ and $\tilde \xi \simeq 1$.
  \item \emph{driving sector}: we introduce three fields,
  $\varphi_0^T$, $\varphi_0^S$, $\xi_0$, that allow to build a non trivial
  superpotential that, once minimized, will ensure the correct vacuum alignment.
\end{enumerate}
On top of $A_4$ we also impose a discrete $Z_3$ symmetry, that is introduced
in order to guarantee that, at the leading order, the charged leptons and the
neutrino sector selectively couple to two different sets of flavons:
$\varphi_T$ for charged leptons and $\varphi_S,\xi,\tilde\xi$ for neutrinos.
This ensures that the vacuum alignment produces the tribimaximal mixing matrix
for the neutrinos~\cite{A4Z3}. The symmetry group also includes a $U(1)_{FN}$
Froggatt-Nielsen factor, that acts only on the RH charged leptons. It is spontaneously
broken by the VEV of a flavon field $\theta$ carrying a negative unit of the Froggatt-Nielsen
charge. By choosing $U(1)_{FN}$ charges $(2,1,0)$ for $(e^c,\mu^c,\tau^c)$,
the mass hierarchy in the charged lepton sector can be reproduced.
Besides the flavor symmetry,
 a continuous $U(1)_R$ $R$-symmetry, containing the usual R-parity of the SUSY
 models is also needed.
 Matter, symmetry-breaking and driving fields have
respectively $R$-charge $1,0$ and 2.
The fields assignments to the $A_4\times Z_3\times U(1)_{FN}$ irreps
are summarized in Table~\ref{tb:rappresentazioni}.
 \begin{table}[h!!]
\begin{center}
\begin{tabular}{|c|c|c|c|c|c|c|c|c|c|c|c|c|c|c|c|c|}
 \hline
 & $L$ & $e^c$ & $\mu^c$ & $\tau^c$ & $N$ & $H_{u,d}$ & $\varphi_T$&
$\varphi_S$&$\xi$&$\tilde\xi$& $\theta$ &$\varphi_0^T$&$\varphi_0^S$&$\xi_0$\\
\hline
$A_4$ & 3&1&1''&1'&3&1&3&3&1&1& 1&3&3&1\\
\hline
$Z_3$&$\omega$&$\omega^2$&$\omega^2$&$\omega^2$&$\omega^2$&1&1&$\omega^2$&$\omega^2$&$\omega^2$& 1&1&$\omega^2$&$\omega^2$\\
\hline
$U(1)_{FN}$ & 0&2&1&0&0&0&0&0&0& 0&-1&0&0&0\\
\hline
$U(1)_{R}$ & 1&1&1&1&1&0&0&0&0&0&0&2&2&2\\
\hline
\end{tabular}\caption{The fields of the $A_4\times Z_3\times U(1)_{FN}$ model and their
 representations.}
\label{tb:rappresentazioni}
\end{center}
\end{table}

The most general superpotential compatible with the representation assignment
of Table~\ref{tb:rappresentazioni} is given by
  $$ w= w_L + w_D$$
  $w_L$ is the leptonic part of the superpotential,
  \begin{eqnarray}\label{w_l}
 w_L &=&y_e \left(\dd\frac{\varphi_T}{\Lambda} L\right)_1 e^c H_d+
y_\mu  \left(\dd\frac{\varphi_T}{\Lambda} L\right)_{1'}\mu^c H_d +
y_\tau  \left(\dd\frac{\varphi_T}{\Lambda}L\right)_{1''}\tau^cH_d\nonumber\\
& +& y \left(LN\right)_1 H_u+\left(x_a {\xi} +
\tilde{x_a}\tilde\xi\right)
\left(NN\right)_1+x_b \left(\varphi_S NN\right)_1+\dots
\end{eqnarray}
 where the dots stand for higher dimensions operators,
 suppressed by additional powers of the cutoff $\Lambda$.
 We note that it is  always possible to redefine the fields
 in a way that $y_e$, $y_\mu$, $y_\tau$ and $y$ are real numbers,
 while $x_a$, $\tilde x_a$ and $x_b$ are in general complex parameters.
 $w_D$ is the ``driving'' part of the superpotential
 \begin{eqnarray}\label{w_d}
 w_D &=&m \left( \varphi_0^T \varphi_T\right)_1+g \left(\varphi_0^T \varphi_T\varphi_T\right)_1\nonumber\\
& +&  g_1 \left(\varphi_0^S \varphi_S\varphi_S\right)_1+g_2 \tilde\xi \left(\varphi_0^S \varphi_S\right)_1+g_3\xi_0\left(\varphi_S\varphi_S\right)_1\nonumber\\
&+&g_4 \xi_0\xi^2+g_5 \xi_0\xi\tilde\xi+g_6\xi_0{\tilde\xi}^2+...\nonumber
\end{eqnarray}
where dots denote higher dimensional contributions.
For a detailed discussion about the minimization of the
superpotential we refer to~\cite{A4Z3}; here we only mention that,
given the previous expression for $w_D$, the VEVs of the
flavon fields at the leading order are
\begin{equation}\label{eq:vacuum_alignment}
 \begin{array}{lll}
 \langle \varphi_T \rangle &=& v_T (1,0,0)\\
 \langle \varphi_S \rangle &=& v_S (1,1,1)\\
 \langle \xi \rangle &=& u\neq 0\\
 \langle \tilde\xi \rangle &=& 0
\end{array}
\end{equation}
where $v_T$, $v_S$ can be expressed in term of $m$, $g$ and $g_i$ ($i=1\dots
6$) while $u$ remains undetermined. We assume that all VEVs in the flavon sector are
of the same order: $v_T\approx v_S\approx u\approx V$. Their value in units of the cutoff
scale $\Lambda$, $V/\Lambda$, corresponds to the parameter $\eta$ of the previous section
\footnote{The precise definition of $\eta$ in this model will be given in section 4.}.

From eq.~(\ref{w_l}) we see that already in the phase when the flavor symmetry
is unbroken, the Yukawa matrix for the singlet neutrinos $Y^0$ is
non-vanishing:
\begin{equation}\label{eq:Yukawa}
{Y^0} =y  \left(
\begin{array}{ccc}
 1&0&0\\
0&0&1\\
0&1&0\\
\end{array}\right).
\end{equation}
In contrast, as long as  $A_4\times Z_3\times U(1)_{FN}$ is unbroken,
the Yukawa matrix for the charged
leptons and the Majorana  neutrinos mass matrix both
vanish. In terms of the expansions eq.~(\ref{eq:expansion}) we thus have
${Y^e}^0=M^0=0$.
After the breaking of the flavor symmetry we get
\begin{equation}\label{eq:leptonMasse}
Y^e= \delta Y^e = \left(
\begin{array}{ccc}
 y_e&0&0\\
0&y_\mu&0\\
0&0&y_\tau\\
\end{array}\right)\frac{v_T}{\Lambda},
\end{equation}
\begin{equation}\label{eq:M}
M= \delta M=  \left(
\begin{array}{ccc}
 a+2b/3&-b/3&-b/3\\
-b/3&2b/3&a-b/3\\
-b/3&a-b/3& 2 b/3
\end{array}
\right),
\end{equation}
where $a=2x_a u$ and  $b=2 x_b v_S$ are complex numbers.
The complex symmetric matrix $M$ is diagonalized by the  transformation
\begin{equation}\label{eq:U}
\hat M= U\, M\, U^T
\end{equation}
where $\hat M$ is a diagonal  with real and positive entries given by
\begin{equation}
\label{eq:RHmasses}
\hat M\equiv \mathrm{diag}(M_1, M_2, M_3)=
\mathrm{diag}(\vert a+ b\vert, \vert a \vert, \vert-a+b\vert).
\end{equation}
The unitary matrix  $U$ can be written as
\begin{equation}\label{eq:UnitaryRH}
    U^T=U_{TB} U_{PH},
\end{equation}
where $U_{TB}$ is the TBM matrix
\begin{equation}\label{eq:UTB}
 U_{TB}=  \left(
\begin{array}{ccc}
\dd\sqrt{\frac{2}{3}} &\ut&0\\ [8pt]
-\us&\ut&-\ud\\  [8pt]
-\us&\ut&\ud
\end{array}
\right)
\end{equation}
and $U_{PH}= \mathrm{diag}(e^{i\alpha_1/2},e^{i\alpha_2/2},e^{i\alpha_3/2})$
is a  matrix of phases, with
$\alpha_1=-\mathrm{arg}(a+b)$, $\alpha_2=-\mathrm{arg}(a)$ and
$\alpha_3=-\mathrm{arg}(-a+b)$.\\

After electroweak symmetry breaking $\langle H_{u,d}\rangle=v_{u,d}$ the mass
matrix for the light neutrinos is obtained from the seesaw formula:
\begin{equation}
  \label{eq:seesaw}
m_\nu= - v_u^2 {Y^0} M^{-1} {Y^0}^T = -v_u^2 y^2 M^{-1}.
\end{equation}
Since from eq.~(\ref{eq:U})
$U^*M^{-1} U^\dagger={\rm diag}(M_1^{-1},M_2^{-1},M_3^{-1},)$,
we obtain for the light neutrinos mass  eigenvalues
\begin{equation}\label{eq:LHmasses}
m_i= \frac{y_\beta^2 v^2}{ M_i} \, ,
\end{equation}
where we defined
\be
y_\beta=y \sin\beta~~~,~~~~~~~\tan\beta=\dd\frac{v_u}{v_d}~~~,
\ee
and $v=\sqrt{v^2_u+v^2_d}\approx 174$ GeV. Notice that these mass relations
correspond to a particular model of sequential dominance \cite{king}.
The neutrino mixing matrix  is given by
\begin{equation}\label{eq:UnitaryLH}
    U_\nu=U^\dagger=U_{TB} U_{PH}^{*}~~~.
\end{equation}
Thus, at first order in the flavor symmetry breaking,
the neutrino mixing matrix is TBM.\\

\subsection{The  neutrino mass spectrum}

\begin{figure}[t!]
\vskip-2mm
\hskip3mm
 \includegraphics[width=7.5cm,height=6cm,angle=0]{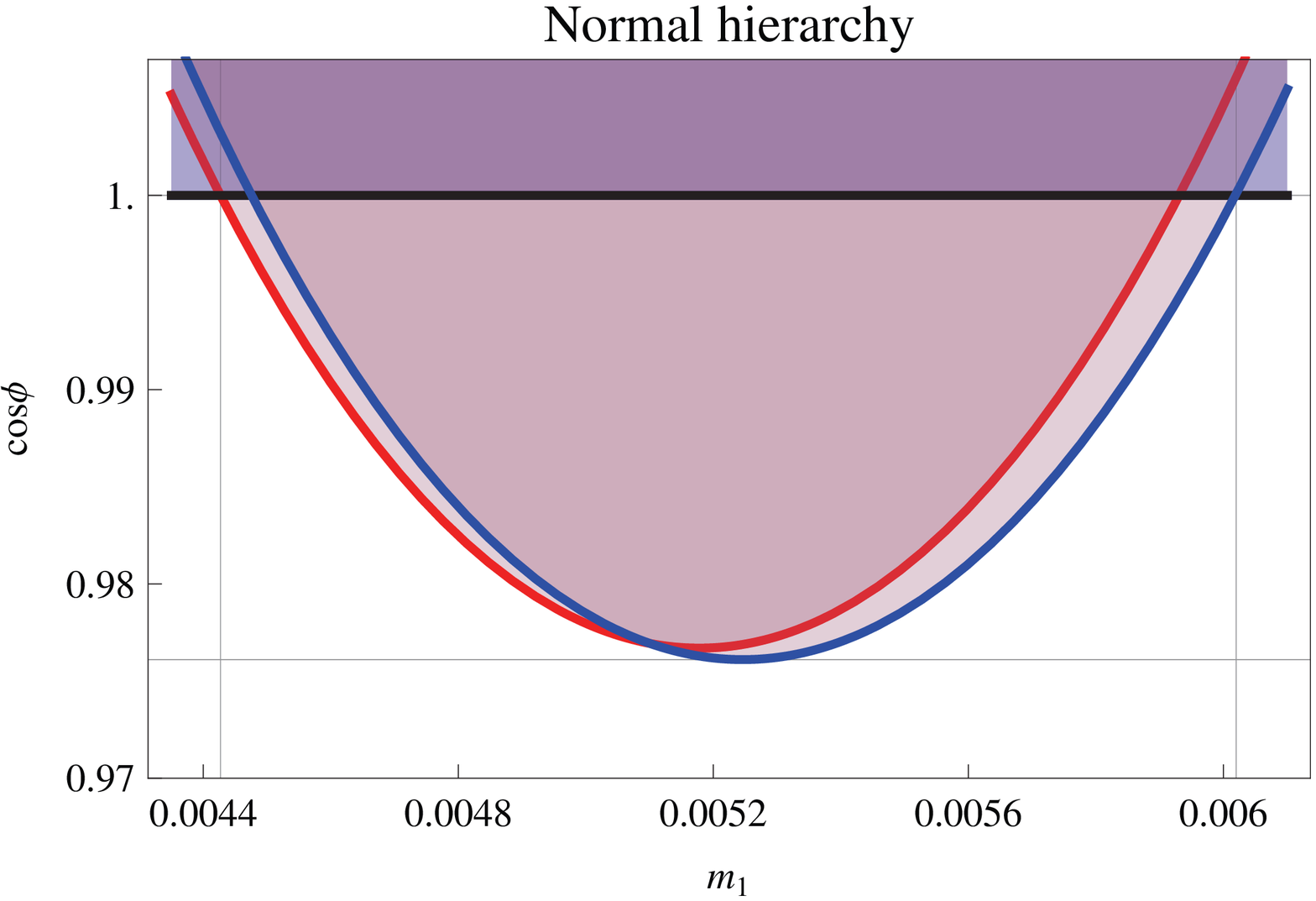}
\hskip5mm
 \includegraphics[width=7.5cm,height=6cm,angle=0]{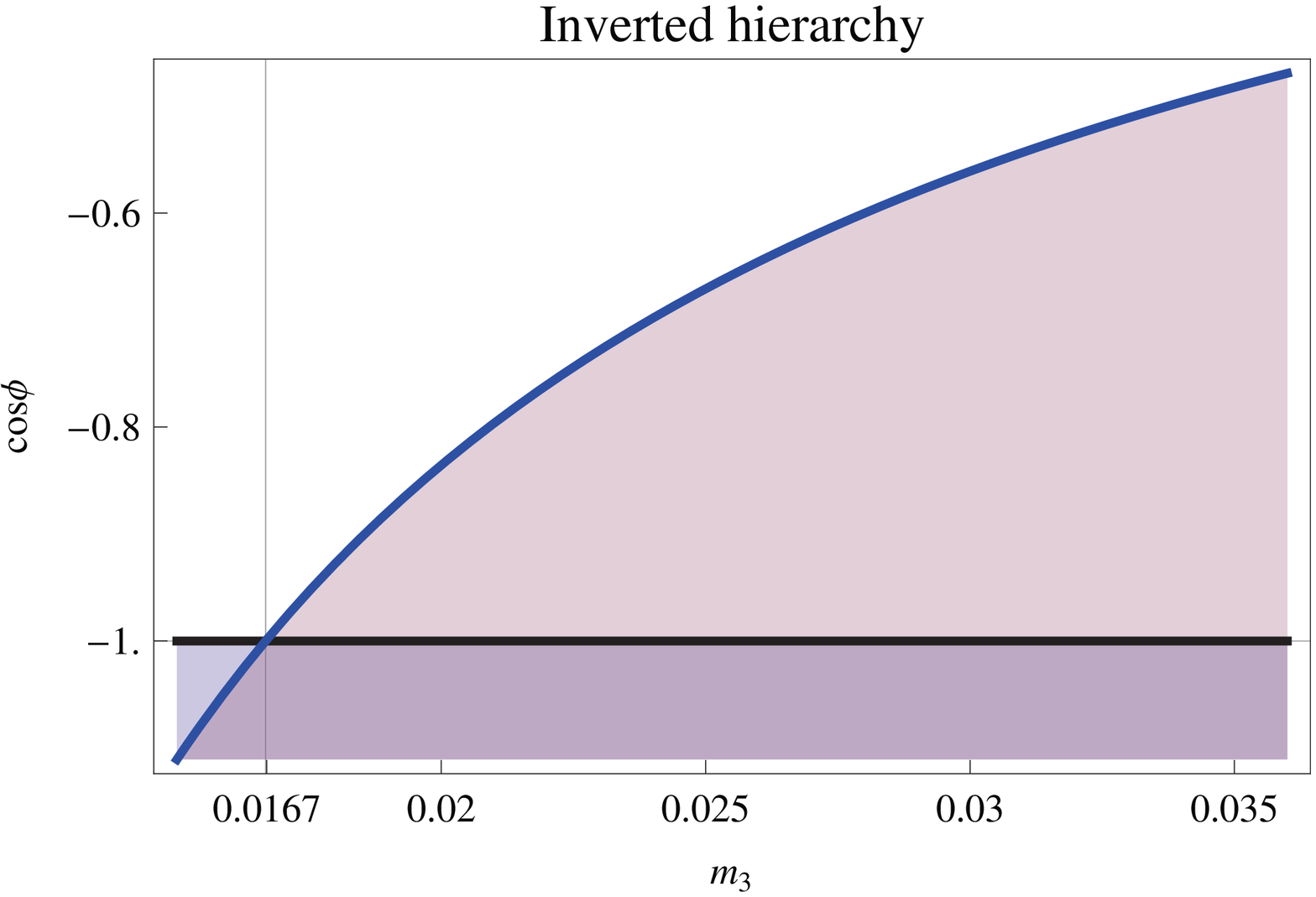}
%
%
\caption[a]{$\cos\varphi$ versus the lightest neutrino mass for the   NO (left) and IO
  (right) cases. For the NO plot the most conservative limits are obtained for
  $\Delta m^2_{atm} = (2.46-0.15)\times 10^{-3}~{\rm eV}^2$
together with
  $\Delta m^2_{sol}= (7.67-0.21)\times 10^{-5}~{\rm eV}^2$
(left red curve),
  and with
$\Delta m^2_{sol}= (7.67+0.22) \times 10^{-5}~{\rm eV}^2$
(right
  blue curve).  For the IO plot the most conservative limit is obtained for
  $\Delta m^2_{atm} = (2.45-0.15)\times 10^{-3}~{\rm eV}^2$
and
$\Delta  m^2_{sol}= (7.67-0.21)\times 10^{-5}~{\rm eV}^2$.
}
\label{fig:cosphi}
\end{figure}

According to~ eq.~(\ref{eq:LHmasses}), in the present approximation
the light neutrino mass spectrum is directly related to the heavy
neutrino masses. These, in turn, can be expressed in terms of just three
independent parameters (cf. \eqn{eq:RHmasses}) that for example can
be chosen to be $|a|=M_2=y_\beta^2 v^2/ m_2$, $|z|$ and $\varphi$. The
latter two are defined according to
\be\label{z}
\frac{b}{a}=|z|e^{i\varphi}\,.
\ee
Experimentally, only two observables related to the spectrum have
been measured. For normal (inverted) hierarchy they are~\cite{maltoni}
\bea
\label{eq:numixingdata}
m_2^2-m_1^2 &\equiv& \  \Delta m^2_{sol}=  (7.67^{+0.22}_{-0.21})\times 10^{-5}~{\rm eV}^2~~~,~~~~~~~\nn\\
|m_3^2-m_1^2 \,(m_2^2)| &\equiv &
\Delta m^2_{atm} =  (2.46\,(2.45)\pm 0.15)\times 10^{-3}~{\rm eV}^2~~~,~~~~~
\label{exp}
\eea
and thus the neutrino mass spectrum is not fully determined.
However, within the present model the value of
a third parameter is bounded  $|\cos\varphi|\leq 1$.
We now show that this condition constrains significatively
the absolute neutrino mass scale. A relation between
the phase $\varphi$ and the neutrino masses is easily derived.
We write
\begin{eqnarray}
  \label{eq:zcosphya32a}
|z|\cos\varphi&=&\frac{1}{4}
\left(\frac{m_2^2}{m^2_1}-\frac{m_2^2}{m_3^2}\right)\,,\\
\label{eq:zcosphya32b}
|z|&=&\sqrt{\frac{1}{2}\left(\frac{m_2^2}{m^2_1}+\frac{m_2^2}{m_3^2}\right)-1}\, ,
\end{eqnarray}
and taking the ratio of the previous two equations we obtain
\be
\label{phase}
\cos\varphi=\frac{\dd\frac{m_2^2}{m_1^2}-\dd\frac{m_2^2}{m_3^2}}
{4\sqrt{\dd\frac{1}{2}\left(\dd\frac{m_2^2}{m_1^2}+\dd\frac{m_2^2}{m_3^2}\right)-1}}~~~.
\ee
This result holds both for normal ordering (NO) and for inverted ordering (IO).
By expressing the heavier masses in~\eqn{phase} in terms of the lightest one
$m_l=m_1\,(m_3)$ for  NO (IO), and  of
$m_{\rm sol} \equiv \sqrt{\Delta m^2_{sol}}\simeq 0.0088\,{\rm eV}$,
$m_{\rm atm} \equiv \sqrt{\Delta m^2_{atm}}\simeq 0.050\,{\rm eV}$
and solving for  the condition
$\left|[\cos\varphi](m_l)\right|\leq 1$, we
obtain, in agreement with \cite{merlo,altamelo}, the following limits for the lightest neutrino mass:
\begin{eqnarray}
  \label{eq:limitNO}
0.0044\> {\rm eV} \leq &m_1& \leq 0.0060\> {\rm eV} \qquad\  {\rm (NO)} \\
  \label{eq:limitIO}
  &m_3 &\geq 0.017\> {\rm eV}  \qquad \ \ \  {\rm (IO)}.
\end{eqnarray}
The function $[\cos\varphi](m_l)$ is plotted in Figure~\ref{fig:cosphi} for
the NO (left) and IO (right).
By expanding $[\cos\varphi](m_l)$  in powers of $r\equiv m_{\rm sol}/m_{\rm atm}$,
we can also derive approximate  analytical expressions
for the limits on $m_l$.  For NO we obtain:
\be
 \frac{\ms}{\sqrt{3}}\left(1-\dd\frac{4\sqrt{3}}{9}\dd\Rmsma
+ ...\right)\le m_1 \le
\frac{\ms}{\sqrt{3}}\left(1+\dd\frac{4\sqrt{3}}{9}\dd\Rmsma
+...\right),
\ee
and for the IO:
\be
m_3\ge \dd\frac{\ma}{2\sqrt{2}}\left(1-\dd\frac{1}{6}\dd\Rmsmasq
+...
\right),
\ee
where the dots represent higher order terms in the expansion in powers of $r$.

For NO we have both a lower and an upper bound on $m_1$, that select a rather
small range for the possible $m_1$ values of width $\sim  \sqrt{3}\,r$
and centered around $\ms/\sqrt{3}$.  Thus, the
neutrino mass spectrum is essentially determined.  From
Fig.~\ref{fig:cosphi} we can also see that the phase $\varphi$ always
remains quite close to zero ($\cos\varphi\lsim \pi/15$), and that both the
upper and lower bounds are saturated for $\varphi=0$.  In the IO case, we only
get a lower bound on $m_3$, that is saturated for $\varphi=\pm\pi$. The
neutrino mass remains unbounded from above, and the phase $\varphi$ is allowed
to vary between $\pi/2$ and $\pi$ or between $-\pi$ and $-\pi/2$, that is in
the ranges where $\cos\varphi$ is negative.

\begin{figure}[t!]
\begin{center}
\psfig{file=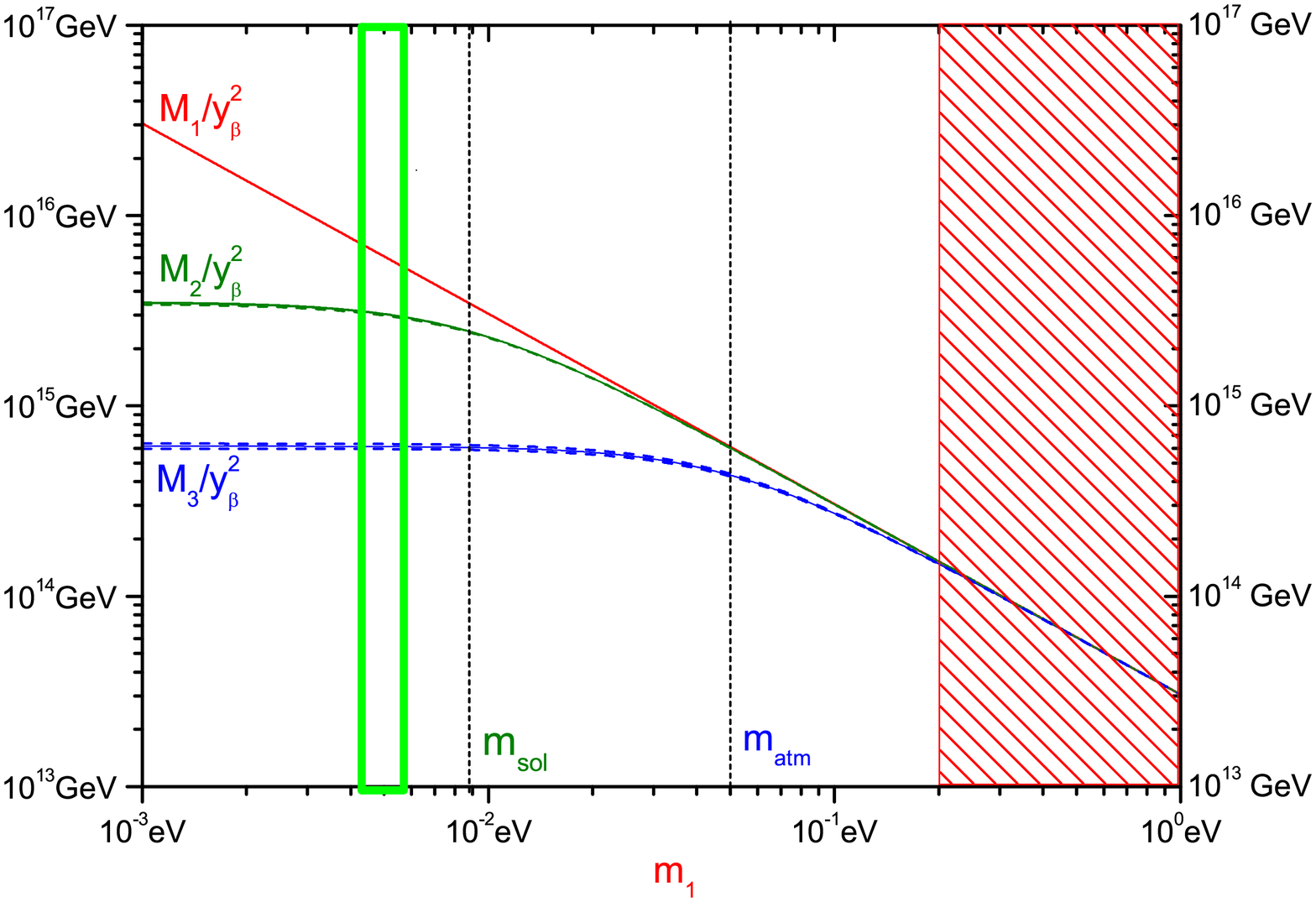,height=6cm,width=7cm,angle=0}
\hspace{-1mm}
\psfig{file=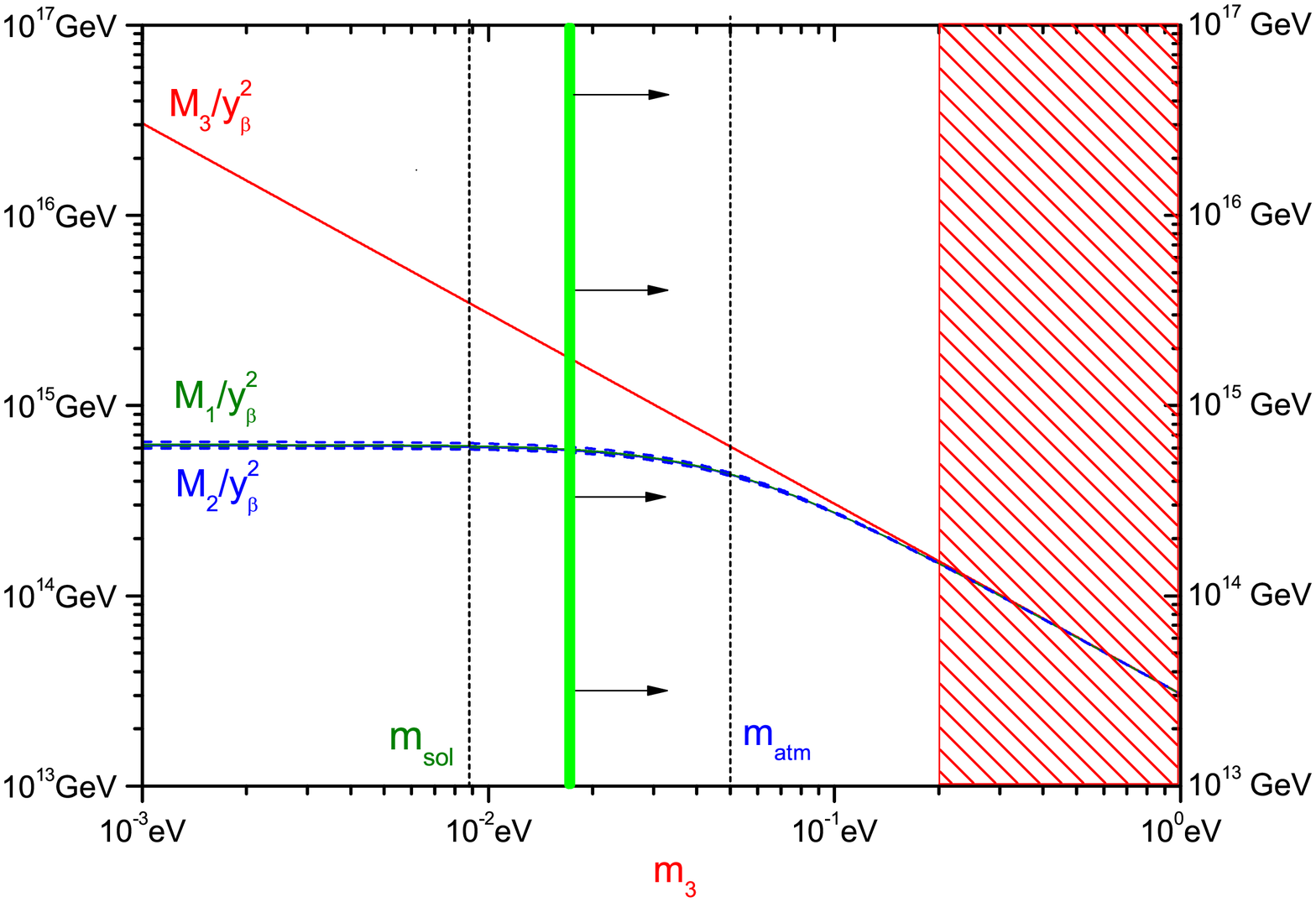,height=6cm,width=7cm,angle=0}
\\
\psfig{file=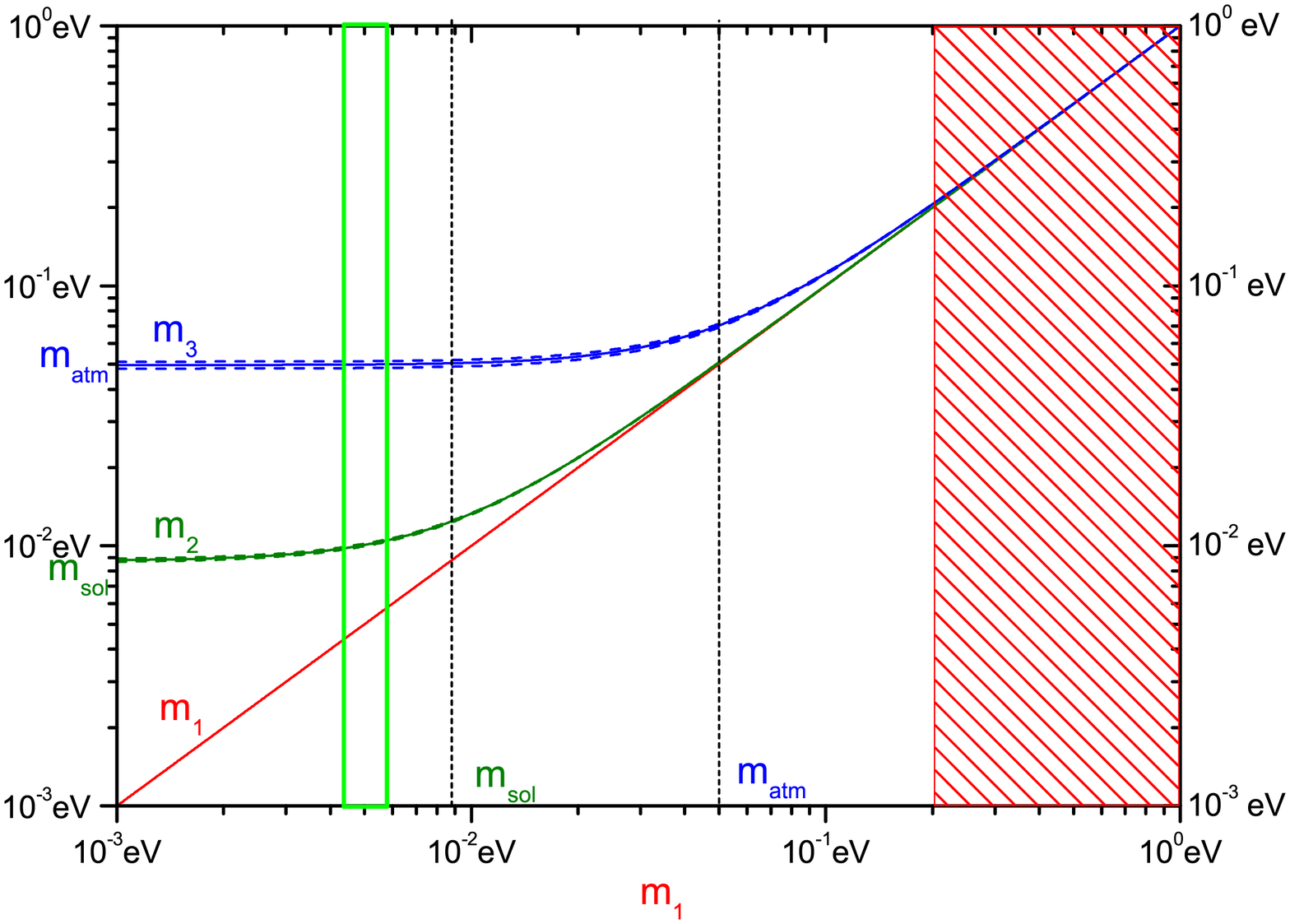,height=6cm,width=7cm,angle=0}
\hspace{-1mm}
\psfig{file=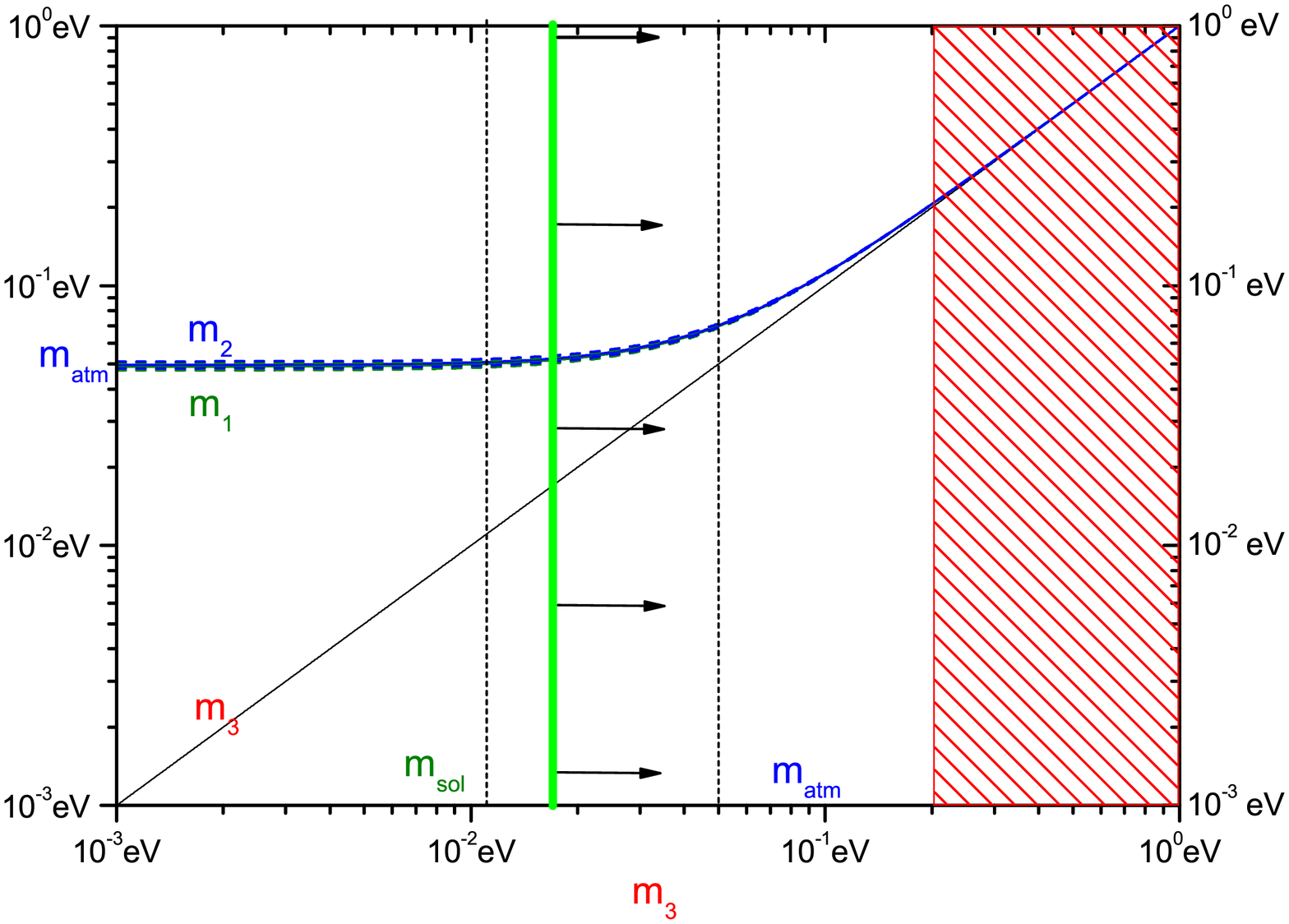,height=6cm,width=7cm,angle=0}
\caption{Plots of the heavy
(upper panels) and light (lower panels)
neutrino masses
divided by $y_\b^2$,
as a function of the lightest
left-handed neutrino mass, according to  eq.~(\ref{eq:LHmasses}).
In the left panels (NO case) the
two vertical green lines show the allowed $m_1$ mass range
as given in~\eqn{eq:limitNO}. In the  right panels (IO case) the vertical green
line shows the lower limit on $m_3$~\eqn{eq:limitIO}. The dashed area
is the region  excluded at $95\%\, {\rm CL}$ by the upper bound
(cf. Eq.~(\ref{upperbound})) obtained by combining the cosmological upper
bound on the sum of neutrino masses with neutrino oscillations data.
}
\label{fig:two}
\end{center}
\end{figure}
The results presented so far are of course approximate since the model gets
corrections when higher dimensional operators are included in the
Lagrangian. The inclusion of higher dimensional operators has also the effect of shifting the VEVs of the flavon fields
from their leading order values, eq. (\ref{eq:vacuum_alignment}). By assuming that the
VEVs of the flavon fields have similar values $v_T\approx v_S\approx u\approx V$,
these corrections modify the leading order approximation by terms of relative order $V/\Lambda$.
The allowed range of $V/\Lambda$ is determined by the requirement that sub-leading corrections which perturb the
leading order result are not too large and by the requirement that the $\tau$ Yukawa coupling $y_\tau$ does not become too large.
The first requirement results in an upper bound on $V/\Lambda$ of about $0.05$, which mainly comes from the fact that the solar
mixing angle should remain in its $1\sigma$ range. The second one gives a lower bound which we estimate as
\be
\dd\frac{V}{\Lambda}\approx\dd\frac{\tan\beta}{y_\tau} \dd\frac{m_\tau}{v} \approx 0.01 \dd\frac{\tan\beta}{y_\tau}~~~.
\label{tanb&u&yt}
\ee
By asking $|y_\tau|<3$ we find a lower limit on $V/\Lambda$
close to the upper bound $0.05$ for $\tan\beta=15$, whereas $\tan\beta=2$ gives as lower limit $V/\Lambda>0.007$.
We choose as maximal range:
\be
0.007< \dd\frac{V}{\Lambda} < 0.05~~~,
\label{ubound}
\ee
which shrinks when $\tan\beta$ is increased from 2 to 15.

In particular the leading order expressions of neutrino masses are modified by terms of
relative order $V/\Lambda$. However, close to $\cos\varphi=\pm 1$, where the bounds
are saturated, the corrections to both the numerator and the
denominator of eq.~(\ref{phase}) remain of relative order $V/\Lambda$, and
thus the bounds in  \eqn{eq:limitNO} and   \eqn{eq:limitIO}
are not significantly affected. For normal hierarchy,
that requires $\cos\varphi$ very close to one
in the full allowed mass range of \eqn{eq:limitNO},
the leading expression for $\cos\varphi$ given in eq.~(\ref{phase})
always remains a good approximation. In the case of inverted
hierarchy, close to $\cos\varphi=-1$,
when $m_3$ is near its lower bound, the corrections are also negligible.
Deviations from eq. (\ref{phase}) can
become significant when $\cos\varphi$ approaches $-0.2$. This happens
for $m_3\approx 0.09$ eV and $m_{1,2}\approx 0.10$ eV, that is when the
spectrum becomes nearly degenerate.

It is interesting to estimate the order of magnitude for the RH
neutrinos masses.  Since no suppression is expected for parameters
that are unrelated to the breaking of the flavor symmetry, we take
$y_\b={\cal O}(1)$.  With this, and using as the light mass scale in the
seesaw equation (\ref{eq:LHmasses}) $m_{\rm atm}\simeq 0.05
\mathrm{eV}$, we obtain
%
%
%
\begin{equation}\label{eq:RHmassEstimate}
    M_i \sim 10^{14\div 15} \mathrm{GeV}\,.
\end{equation}
A detailed summary of the neutrino mass relations and bounds is given in
figure~\ref{fig:two}, where we have plotted the three light and the three
heavy neutrino masses, taking into account both the information from neutrino
oscillations data (cf. eq.~(\ref{eq:numixingdata})) and the seesaw relations
eq.~(\ref{eq:LHmasses}). In this way, at leading order, all six masses
can be expressed as a function of just one independent parameter, that can be
conveniently chosen to be the lightest left-handed neutrino mass $m_l$.
Of course there is also a dependence on the neutrino mass ordering,
namely if  $m_l=m_1$ or $m_l=m_3$. In the figures we also display
the cosmological upper bound on $m_l$
\be\label{upperbound}
m_l < 0.2\,{\rm eV} \;\; (95\%\, {\rm CL}) \, ,
\ee
that follows from the  WMAP5 \cite{WMAP5} upper bound on the
sum of the three neutrino masses combined with  the
constraints  from mass squared differences from
oscillations data (cf. eq.~(\ref{eq:numixingdata})).

\subsection{Neutrinoless double-$\beta$  decay}

In the approximation of neglecting terms of higher order in the
symmetry breaking, as well as RGE running effects from the high scale
to the eV scale, we can straightforwardly obtain predictions for the
$0\nu2\beta$ decay parameter $|m_{ee}|$ for both the NO and IO cases.
The $0\nu2\beta$ decay parameter is defined as
\begin{equation}  \label{eq:mee}
  |m_{ee}|= \left|\sum_i (U_\nu)_{ei}^2  m_i\right|  \,,
\end{equation}
and corresponds to  the (11) entry in the neutrino mass matrix
$m_\nu$ in \eqn{eq:seesaw}:
\begin{equation}
\label{eq:mnu11}
  |m_{ee}|=|(m_\nu)_{11}|
= v_u^2\, y^2\,  \left|\frac{2}{3(a+b)}+\frac{1}{3a}\right|\,.
\end{equation}
In terms of physical neutrino masses,
and to the order we are working here ($(U_\nu)_{13}=0$)
\eqn{eq:mee} assumes the particularly simple form:
\begin{equation}
  \label{eq:mee2}
  |m_{ee}|=
\left|\frac{2}{3}m_1 e^{i\alpha_1}+ \frac{1}{3} m_2 e^{i\alpha_2} \right|\,,
\end{equation}
and thus depends only  the phase difference $\alpha_{2}-\alpha_{1}$,
where $\alpha_i$  are the phases of the diagonal matrix $U_{PH}$
defined below eq.~(\ref{eq:UTB}).

In order to compute the allowed range of $|m_{ee}|$ when the lightest
neutrino mass is allowed to vary in the allowed region \eqn{eq:limitNO}
(for NO) or \eqn{eq:limitIO} (for IO), it is more convenient to express
$|m_{ee}|$ directly in terms of the phase $\cos\phi$ in \eqn{phase}.
This can be done more easily by using \eqn{eq:mnu11}, and yields:
\begin{equation}
|m_{ee}|=\frac{m_2}{3}\sqrt{1+4\>\frac{2+|z|\cos\phi}{1+|z|^2+2|z|\cos\phi}}\,,
\end{equation}
that holds for both NO and IO.

Following \cite{Feruglio:2002af} we  parameterize the forecast
sensitivity of future $0\nu2\beta$ experiments as
\begin{equation}
\label{eq:forecast}
|m_{ee}| \to 10\, h\> {\rm meV},
\end{equation}
where $h=0.6\div 2.8$ parameterizes the theoretical uncertainty related
to different nuclear matrix elements calculations.  We then confront
the prediction of the $A_4\times Z_3\times U(1)_{FN}$ model with \eqn{eq:forecast}.

%
\begin{figure}[t!]
\vskip-2mm
\hskip3mm
 \includegraphics[width=7.5cm,height=6cm,angle=0]{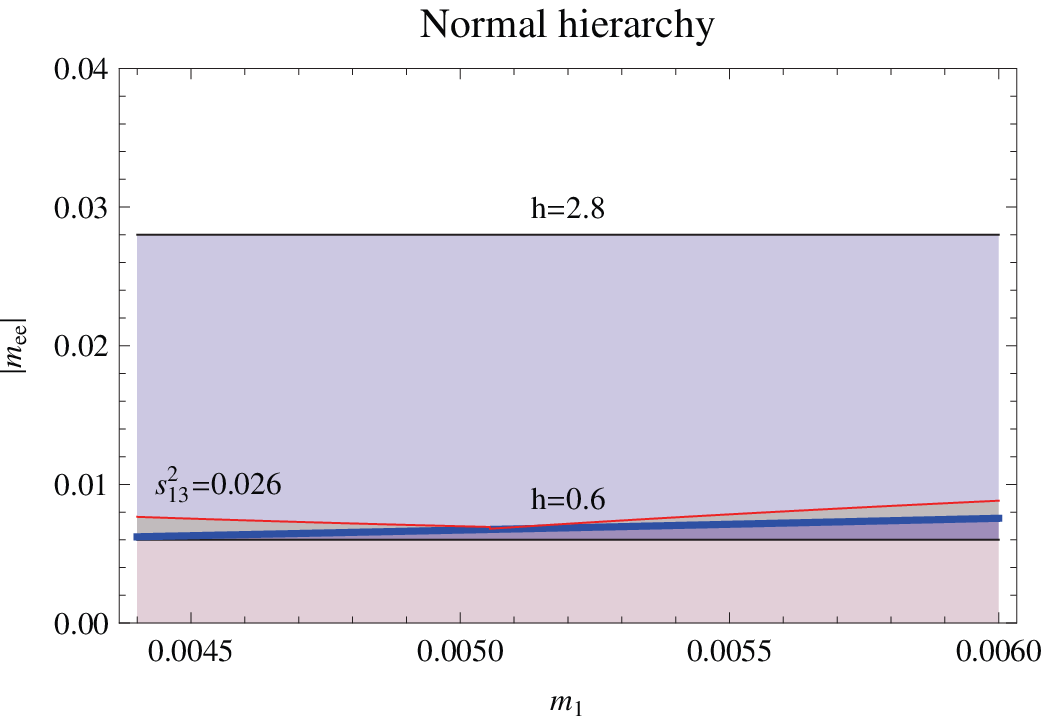}
\hskip5mm
 \includegraphics[width=7.5cm,height=6cm,angle=0]{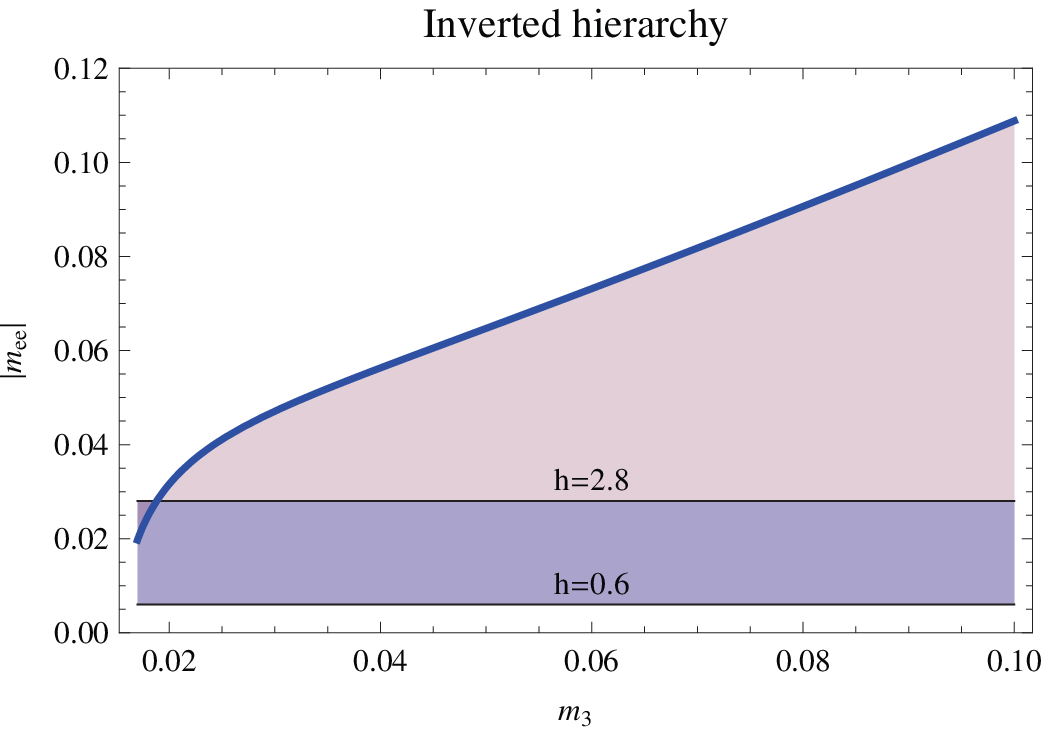}
%
%
\caption[a]{$|m_{ee}|$ versus the lightest neutrino mass.
Left panel:  the   NO case. The thick blue line depict
 the value of $|m_{ee}|$ as a function of $m_1$. The thin red line
illustrates the size of the corrections that could arise
for $\sin^2 \theta_{13} \approx 0.026$.
Right panel: $|m_{ee}|$ as a function of $m_3$ in the IO case.
In both cases the  dark shaded horizontal region depicts the
projected sensitivity of future $0\nu2\beta$ experiments.
}
\label{fig:betadecay}

\end{figure}

For the NO case we have $m_l=m_1$ and, according to \eqn{eq:limitNO},
$m_l$ can vary between $0.0044\,$eV and $0.0060\,$eV.  The results for
$|m_{ee}|$ in this case are depicted with the thick blue line in the
left panel of figure~\ref{fig:betadecay}.  The dark horizontal region
corresponds to the expected sensitivity of future $0\nu2\beta$
experiments according to \eqn{eq:forecast}.  In the leading order
approximation only $m_1$ and $m_2$ contribute to $|m_{ee}|$ (see
\eqn{eq:mee2}).  In the NO case these are the two smaller masses, and
it is then reasonable to ask to what extent effects related to the
largest mass $m_3$, that also contributes to $|m_{ee}|$ when $\sin^2
\theta_{13}\neq 0$, can affect these results. By inserting in $U_\nu$
the maximum value $\sin^2 \theta_{13}\approx 0.026$ allowed at
1$\sigma$ by present data~\cite{maltoni}, with the value of $m_3$
fixed in terms of $m_1$ and the mass squared differences, we obtain
the values of $|m_{ee}|$ depicted in figure~\ref{fig:betadecay} (left
panel) with the thin red line.  Of course, the procedure of setting
$(U_\nu)_{13}\neq 0$ by hand, and not as a result of the inclusion of
higher order terms, can just give a feeling of the possible effects of
a non-vanishing $\sin^2 \theta_{13}$, but does not correspond in any
way to an improved prediction, that would require a consistent
treatment of all higher order effects.

From the left panel in figure~\ref{fig:betadecay} we can then conclude
that for all the allowed values of $m_l$, even in the most optimistic
situation, $|m_{ee}|$ remains at best marginally in the reach of the
future experiments sensitivity. We also see that higher order
contributions $\propto m_3\cdot \sin^2 \theta_{13} $ are not expected
to change this conclusion.

For the IO we have $m_l=m_3$. A plot of $|m_{ee}|$ as a function of
$m_3> 0.017\,$eV (see~\eqn{eq:limitIO}) is depicted with a thick blue
line in the right panel of figure \ref{fig:betadecay}.  We see that in
the IO case, $|m_{ee}|$ remains well above the sensitivity of future
experiments, except in the most pessimistic situation ($h=2.8$) and
when the lower bound is saturated ($m_3\sim 0.017\,$eV). In this
situation the interference of the two contributions in \eqn{eq:mee2}
is maximally destructive, yielding $|m_{ee}|\simeq m_{\rm atm}/3$.
The results plotted in figure~\ref{fig:betadecay} (right panel) will
not be significatively affected by contributions $\propto m_3\cdot
\sin^2 \theta_{13}$, that for IO are surely negligible, or by other
higher order effects.  Thus, the $A_4\times Z_3\times U(1)_{FN}$ model predicts that
if the neutrino ordering is inverted, a $0\nu2\beta$ decay signal
quite likely will be observed in the next future. Detection of
$0\nu2\beta$ decay is guaranteed if new theoretical computations will
establish that $h\lsim 1.7$.

\section{Leptogenesis in the $A_4\times Z_3\times U(1)_{FN}$ model}

In section~\ref{sec:irreducible} we have shown that an important
consequence of non-Abelian flavor symmetries with the $N_i$ assigned
to an irreducible representation of the symmetry group, is that the
leptogenesis CP asymmetries vanish at leading order. Their size is
thus determined by the size of the flavor symmetry breaking parameters
$\eta$.  In the $A_4\times Z_3\times U(1)_{FN}$ the heavy Majorana
masses are bounded from below by $M_i/y_\beta^2 \gsim 1.5\times
10^{14}\,$GeV (see upper right panel in figure \ref{fig:two}), and
thus are generally quite large.  Therefore, the condition $M_i \gtrsim
10^{12}\,(1+\tan^2\beta)\,$GeV, that ensures that leptogenesis occurs
in the unflavored regime, is generally matched by all the $N_i$ for
natural values of the parameters (e.g. $\tan\beta \lsim 10$ and
$y\gsim 0.5$ (NO) or $y\gsim 0.8$ (IO)). In this regime, the relevant
quantities that have to be considered are the total CP asymmetries
$\varepsilon_i$, and since these asymmetries are ${\cal O}(\eta^2)$
(see \eqn{eq:Oeta2}), the requirement of successful leptogenesis can
provide hints on the minimum size of the typical symmetry breaking
effects.  The important point here is that in general the symmetry
breaking parameters are also related to some observables that are
measurable at low energy. For example, in models that predict
neutrinos tribimaximal mixing (TBM) (as is the case of our $A_4\times
Z_3\times U(1)_{FN}$ model) both the value of $\theta_{13}$ and the
deviation from maximal angle of $\theta_{23}$ are related to $\eta$
and, by studying leptogenesis, one can infer preferred ranges of
values for these parameters.  This is a new type of
low-energy/high-energy connection (see however~\cite{Jenkins:2008rb,Lin})
and constitutes the main motivation of the following analysis.

\subsection{The leptogenesis CP asymmetries at subleading order}

Since in our $A_4$ model the heavy Majorana neutrinos belong to an
irreducible representation of the flavor group, independently of the
particular basis we have that $ {\cal Y}^0\equiv {Y}^{0\dagger} {Y^0}
= y^2\, \mathbb{I}$.  Thus, at leading order, all the CP asymmetries
vanish.
When the leading symmetry breaking terms are
introduced in the Lagrangian,  new complex parameters are
generated that give rise to non-vanishing CP asymmetries
(see~eqs.~(\ref{eq:Oeta}) and (\ref{eq:Oeta2})).

All the possible leading order corrections to the $A_4\times Z_3\times U(1)_{FN}$
 model have been listed in \cite{A4Z3}.  However, as long as
the CP asymmetry are concerned, the relevant terms are only
two~\cite{Jenkins:2008rb}, and both represent corrections to the
Majorana neutrinos Yukawa matrix:
\begin{equation}
\dd\frac{y_1}{\Lambda} \left( \varphi_T \left( \ell N\right)_S\right)H_u,
\qquad\qquad
 \dd\frac{y_2}{\Lambda}\left( \varphi_T \left(\ell N \right)_A\right)H_u\,.
\end{equation}
Using the product rules of eqs.~(\ref{eq:uno}-\ref{eq:tre_A}), the Yukawa
matrix  now reads:
\begin{equation}\label{eq:Ysubleading}
 Y=Y^0+\delta Y=\left(
\begin{array}{ccc}
 y+2d & 0 & 0\\
 0 & 0 & y -d +c\\
 0 & y - d - c & 0 \\
\end{array}
\right)
\end{equation}
where we have introduced the two complex numbers $c,d$ ($|c|,|d|\ll y$)
defined as
\begin{equation}
c\equiv\frac{y_2 v_T}{2\Lambda}, \qquad\qquad
d \equiv\frac{y_1 v_T}{3\Lambda}\,.
\end{equation}
Assuming, as is reasonable to do, that $y_1$ and $y_2$ are
numbers of the same order, we can parameterize the  size of the
symmetry breaking effects with a single hierarchical parameter $\eta$,
defining for example
\be
 \eta={\rm Re}(d) \hspace{10mm} \mbox{\rm and} \hspace{10mm} \rho= {{\rm Re}(c)\over {\rm Re}(d)} \, ,
\ee
where $\rho={\cal O}(1)$. Then in the hatted basis
the matrix $\hat{\cal Y}={\hat Y}^\dagger {\hat Y}$
becomes
%
\begin{equation}\label{calY}
\begin{array}{lll}
   \hat{\cal Y}& =& U_{PH}^{-1}  U_{TB}^T  {\cal Y}~  U_{TB} U_{PH}\\
   &&\\
 &=&y\cdot  \left(
\begin{array}{ccc}
  y +2\,\eta
& 2\sqrt{2}\,\eta\, e^{i\alpha_{21}}
& - \dd\frac{2}{\sqrt{3}}\, \r\,\eta\, e^{i\alpha_{31}}
\\
 2\sqrt{2}\,\eta\, e^{i\alpha_{12}}
& y
& 2\sqrt{\dd\frac23}\, \r\,\eta\, e^{i\alpha_{32}}
\\
-\dd\frac{2}{\sqrt 3}\, \r\,\eta\, e^{i\alpha_{13}}
&2\sqrt{\frac23}\,\r\,\eta\, e^{i\alpha_{23}}
&y-2\,\eta
\\
\end{array}\right)\,.
\end{array}
\end{equation}
Since $\hat{\cal Y}$ is a physical observable (it can be measured in
principle by measuring the CP asymmetries) it depends only on phase
differences $\alpha_{ij}\equiv (\alpha_i-\alpha_j)/2$.

Recalling now the expressions of the CP asymmetries given
in~\eqn{eq:CPasymm} in terms of $\hat{\cal Y}$ and of the functions
$f_{ij}$ (see \eqn{eq:fgij}) we can write:
\begin{eqnarray}
\label{asimmetria_CP}
 \varepsilon_1&=&\frac{\eta^2}{8\pi}\left(8\,
\sin\left(2\alpha_{21}\right) f_{12}+\frac 43 \,
\r^2\,\sin\left(2\alpha_{31}\right)f_{13}\right)\nonumber\\
 \varepsilon_2 &=& \frac{\eta^2}{8\pi}
\left(-8\,\sin\left(2\alpha_{21}\right)f_{21}+
\frac 83\,\r^2\,\sin\left(2\alpha_{32}\right)f_{23}\right)\nonumber\\
 \varepsilon_3 &=& \frac{\eta^2}{8\pi}
\left(-\frac 43\, \r^2\,\sin\left(2 \alpha_{31} \right)
f_{31}-\frac 83\, \r^2\,\sin\left(2\alpha_{32}\right)f_{32}\right)\,.
\end{eqnarray}
Starting from $2\,\alpha_{21}=-\mathrm{arg}(a)+\mathrm{arg}(a+b)$
and $2\,\alpha_{32}=\mathrm{arg}(a)-\mathrm{arg}(-a+b)$, one finds
the following relations
\bea\label{eq:a21a32}
\nonumber
\sin(2\,\alpha_{21}) & = & {|z|\,\sin\varphi \over \sqrt{1+2\,|z|\,\cos\varphi+|z|^2}} \, , \\ \nonumber
\sin(2\,\alpha_{32}) & = & -{|z|\,\sin\varphi \over \sqrt{1-2\,|z|\,\cos\varphi+|z|^2}} \, ,\\
\sin(2\,\alpha_{31}) & = & -{2\,|z|\,\sin\varphi \over \sqrt{(|z|^2-1)^2+4\,|z|^2\,\sin^2\varphi}} \, .
\end{eqnarray}
We recall that, after
expressing the heavier neutrino masses in terms of $m_{\rm sol}$,
$m_{\rm atm}$ and $m_l$, one has that $\sin(2\alpha_{21})$,
$\sin(2\alpha_{32})$ and $\sin(2\alpha_{31})$ in~\eqn{eq:a21a32} are a function of $m_l$ only.
Moreover, given that the expressions for the $\varepsilon_i$ depend, through the functions $f_{ij}$,
only on ratios of the heavy Majorana masses $M_i$, and that through
the seesaw formula~\eqn{eq:seesaw} these ratios are directly related
to the ratios of light neutrino masses $m_i$, we can conclude that the
CP-asymmetries in~\eqn{asimmetria_CP} depend only on $m_l$, on the
non-hierarchical parameter $\rho$, and on the parameter $\eta$ that
quantifies the flavor symmetry breaking effects.\\

\subsection{Analytic approximations}

 In this Section we present simple analytical formulae to estimate the
 matter-antimatter asymmetry produced via leptogenesis within the
 $A_4\times Z_3\times U(1)_{FN}$ model.  The presence of the heaviest
 RH neutrino makes possible for the $C\!P$ asymmetry $\ve_2$ of the
 next-to-lightest RH neutrino to be unsuppressed
 (cf. (\ref{asimmetria_CP})) compared to the $C\!P$ asymmetry of the
 lightest RH neutrino, even when the RH neutrino spectrum is strongly
 hierarchical~\cite{geometry}. Furthermore, since in our case the
 heavy neutrino spectrum is only mildly hierarchical, even the $C\!P$
 asymmetry of the heaviest RH neutrino is not particularly suppressed.
 Therefore, all the three contributions to the final asymmetry have to
 be taken into account since in the presence of compensating effects
 from reduced wash-outs the $B-L$ asymmetry generated in the decays of
 the heaviest RH neutrinos could become comparable to the one
 generated in the decays of the two lighter RH neutrinos.  This
 situation is indeed realized in the NO case.

For our estimates, we adopt the following simplifications:

\begin{itemize}

\item For the reasons explained in the beginning of the Section, we assume
that leptogenesis occurs in the unflavored regime.

\item The $A_4\times Z_3\times U(1)_{FN}$ model must be supersymmetric, since the
  vacuum alignment conditions \eqn{eq:vacuum_alignment} are fulfilled
  within a supersymmetric framework~\cite{A4Z3}.  Our results are
  instead obtained neglecting all supersymmetric partners effects.
  This underestimates the resulting $B-L$ asymmetry by a factor
  $\approx \sqrt{2}$ (see ref.~\cite{proc} and \cite{Davidson:2008bu} Sec. 10).

\item The value of the final $B-L$ asymmetry that we estimate is
  obtained by summing up the asymmetries generated in the decays of
  the three heavy neutrinos~\cite{bcst,N2review2} but neglecting the
  wash-out of the asymmetry due to the inverse processes of the
  lighter RH neutrinos \cite{egnn,N2review1}.  This can be done
  because, neglecting ${\cal O} (\eta^2)$ terms, the leptons produced
  in the decays of the three RH neutrinos are orthogonal to each other:
  \be
\label{Pab}
|\langle L_i|L_j \rangle|^2 =
\frac{|\hcY_{ji}|^2}{|\hcY_{jj}\hcY_{ii}|}= \delta_{ij} + {\cal O}
(\eta^2) \, .
\ee
Doing this we are slightly overestimating the final asymmetry since we
neglect the fact that the three lepton states $|L_i\rangle =
(\hcY_{ii})^{-1/2}\,\sum_\alpha \hat Y^*_{\alpha i} |L_\alpha\rangle$
produced in $N_i$ decays are not exactly orthogonal one to the other,
and thus part of the asymmetry produced by the heavier states gets
washed out by the lighter RH neutrinos interactions.
These effects could be taken into account
in a straightforward way following the procedure explained in
in~\cite{egnn}.
However, neglecting the wash-out from lighter RH
neutrinos is certainly consistent with the order of our approximation
since, e.g.  relative ${\cal O} (\eta)$ corrections to the CP
asymmetries, that could produce even larger effects, are also
neglected. With this approximation we also do not have to worry
about complications coming from an overlap between decays and inverse
decays that occur when the RH neutrino mass spectrum is not strongly
hierarchical, and the leptons produced in the decays of the three RH
neutrinos are not orthogonal to each other~\cite{beyond}.

\item We  neglect subleading leptogenesis effects like
$\D L=1$ scatterings  \cite{luty,plumacher,giudice,plumacher2}
and CP violation in scatterings~\cite{Nardi:2007jp},
thermal corrections \cite{giudice},
spectator processes \cite{buchplum,Nardi:2005hs},
departure from kinetic equilibrium \cite{hannestad}.
In the strong wash-out regime that is the relevant one
for our model, these effects give corrections at most at
the level of $\sim 50\% $.


\end{itemize}

With these approximations, the $B-L$ asymmetry can be estimated by
solving the following three independent pairs of Boltzmann equations

\begin{eqnarray}
{dN_{N_i}\over dz_i} & = &
-D_i\,(N_{N_i}-N_{N_i}^{\rm eq}) \;,
\hspace{10mm} (i=1,2,3)\label{dlg1}
\\
{dN^{(i)}_{B-L}\over dz_i} & = &
\varepsilon_i\,D_i\,(N_{N_i}-N_{N_i}^{\rm eq})-
N_{B-L}^{(i)}\,[W_i(z_i)+\Delta W_i(z_i)] \, ,
\label{dlg2a}
\end{eqnarray}
where $z_i \equiv M_i/T$. We indicated with $N_X$
any particle number or asymmetry $X$ calculated in a portion of co-moving
volume containing one heavy neutrino in ultra-relativistic thermal equilibrium,
so that $N^{\rm eq}_{N_i}(T\gg M_i)=1$.
With this convention the predicted baryon-to-photon ratio $\eta_B$ is
related to the final value of the $B-L$ asymmetry
$N_{B-L}$ by the relation
\be\label{etaB}
\eta_B=a_{\rm sph} {N_{B-L}^{\rm f}\over N_{\g}^{\rm rec}}\simeq 0.96\times
10^{-2}\, N_{B-L}\, ,
\ee
where $N_{\g}^{\rm rec}\simeq 37$, and $a_{\rm sph}=28/79$.
The decay factors are given by
\be
D_i \equiv {\G_{{\rm D},i}\over H\,z_i}=K_i\,z_i\,
\left\langle {1\over\gamma_i} \right\rangle   \, .
\ee
Moreover, indicating with $g_{\star}=g_{SM}=106.75$ the total number of degrees of freedom
and with $M_{\rm Pl}=1.22\,\times\, 10^{19}\,{\rm GeV}$ the Planck mass,
the expansion rate can be expressed as
\begin{equation}
H(z_i)= \sqrt{8\,\pi^3\,g_{\star}\over 90} {M_i^2\over M_{\rm Pl}}\,{1\over z_i^{2}}
\simeq 1.66\,\sqrt{g_{\star}}\,{M_i^2\over M_{\rm Pl}}\,{1\over z_i^{2}} \, .
\end{equation}
The total decay rates, $\G_{{\rm D},i} = (\G_i+\bar{\G}_i)\,\langle
1/\gamma_i\rangle$, are the product of the decay widths times the
thermally averaged dilation factors that can be expressed in terms of
the ratio of the modified Bessel functions, such that $\langle
1/\gamma_i\rangle={\cal K}_1(z_i)/ {\cal K}_2(z_i)$ .  The equilibrium
abundance and its rate can be also expressed in terms of the modified
Bessel functions:
\be
 N_{N_i}^{\rm eq}(z_i)= {1\over 2}\,z_i^2\,{\cal K}_2 (z_i) \;\; ,
\hspace{10mm}
{dN_{N_i}^{\rm eq}\over dz_i} = -{1\over 2}\,z_i^2\,{\cal K}_i (z_i) \, .
\ee
Introducing the effective washout parameters~\cite{plumacher}
\be
\widetilde{m}_i=v^2\frac{\mathcal{Y}_{ii}}{M_i}
\ee
and the equilibrium neutrino mass \cite{orloff,nubounds}
\be
m_\star=\frac{16 \pi^{5/2} \sqrt{g_\star}}{3 \sqrt{5}}\frac{v^2}{M_{Pl}}\simeq 1.08
\times 10^{-3}~\mathrm{eV} \, ,
\ee
the decay parameters can be expressed as
\be
K_i\equiv {\G_i+\bar{\G}_i\over H(z_i=1)} = {\widetilde{m}_i\over m_*} \, .
\ee
In our case, from the Eq.~(\ref{calY}) one can verify that $\widetilde{m}_i\simeq m_i$
and therefore there is a very simple relation between neutrino masses and decay parameters.

After proper subtraction of the resonant contribution from
$\Delta L=2$ processes~\cite{dolgov}, the inverse decay
washout terms are  given by
\be\label{WID}
W_i(z_i) =
{1\over 4}\,K_i\,{\cal K}_1(z_i)\,z_i^3 \, .
\ee
The wash-out term $\D W_i(z_i)$ is the non-resonant contribution to the
wash-out coming from $\D L=2$ processes and can be written as
\be
\Delta W_i(z_i) \simeq {\a \over z_i^2}\,M_i\,\widetilde{m}_i^2 \, ,
\ee
where
\be
\a={3\,\sqrt{5}\,M_{\rm Pl}\over
4\,\zeta(3)\,\pi^{9/2}\,v^4\,\sqrt{g_{\star}}} \, .
\ee
The $B-L$ asymmetry  produced from $N_i$-decays
can then be estimated as~\cite{pedestrians}
\be
\label{NBmLun}
N^{(i)}_{B-L}= \,\ve_i\,\kappa(K_i)\,e^{-{\a\over z_B(K_i)}\,M_i\,\widetilde{m}_i^2} ,
\ee
where  $\k(K_i)$ accounts for the wash-out from inverse processes and is
approximately given by
\be
\k(K_i)={2\over K_i\,z_B(K_i)}\,
\left(1-e^{-{K_i\,z_B(K_i)\over 2}}\right) \, .
\ee
The quantity
\be
z_B(K_i) \simeq 2+4\,{K_i}^{0.13}\,e^{-{2.5\over K_i}}
\ee
gives the approximate the value of $z_i$ around which the final
asymmetry from $N_i$-decays is dominantly produced.

The exponential factor in eq.~(\ref{NBmLun}) accounts for the wash-out
from $\D L=2$ processes. One can notice that the two wash-out
contributions factorize. Notice also that the $\Delta L=2$ processes
suppression is relevant only for $M_i \gtrsim 10^{14}\,{\rm
  GeV}\,(0.05\,{\rm eV}/m_i)^2$.

Finally, using  \eqn{NBmLun} and \eqn{etaB},
the baryon to photon number ratio can be written as
\be\label{etaB2}
\eta_B =\sum_i \eta_{i} \simeq
0.96\times 10^{-2}\ \sum_i\, \ve_i\,\k(K_i)\,\,e^{-{\a\over z_B(K_i)}\,M_i\,m_i^2}.
\ee
This theoretical prediction has to be compared with the observed
value from WMAP data \cite{WMAP5}
\be\label{etaBobs}
\eta_B^{\rm CMB} = (6.2 \pm 0.15)\times 10^{-10} \, .
\ee

\subsection{Results}

We now describe the results for $\eta_B$,
separating the discussion for the NO and the IO case.

\subsubsection{Normal Ordering}

\begin{figure}
\begin{center}
\psfig{file=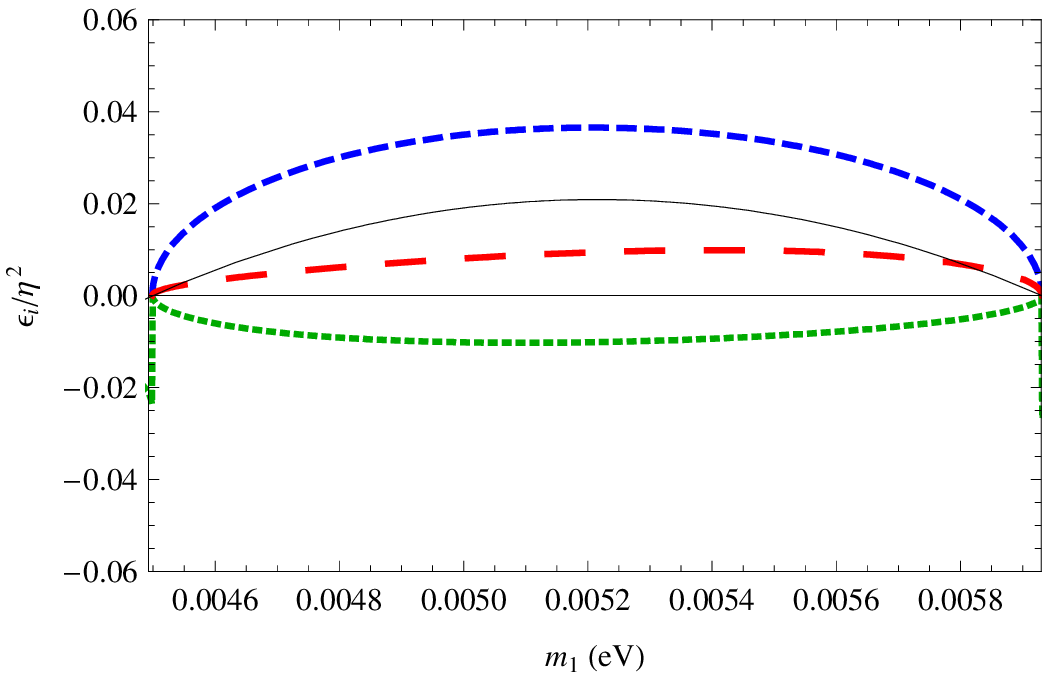,height=4.2cm,width=5.2cm,angle=0}
\hspace{-1mm}
\psfig{file=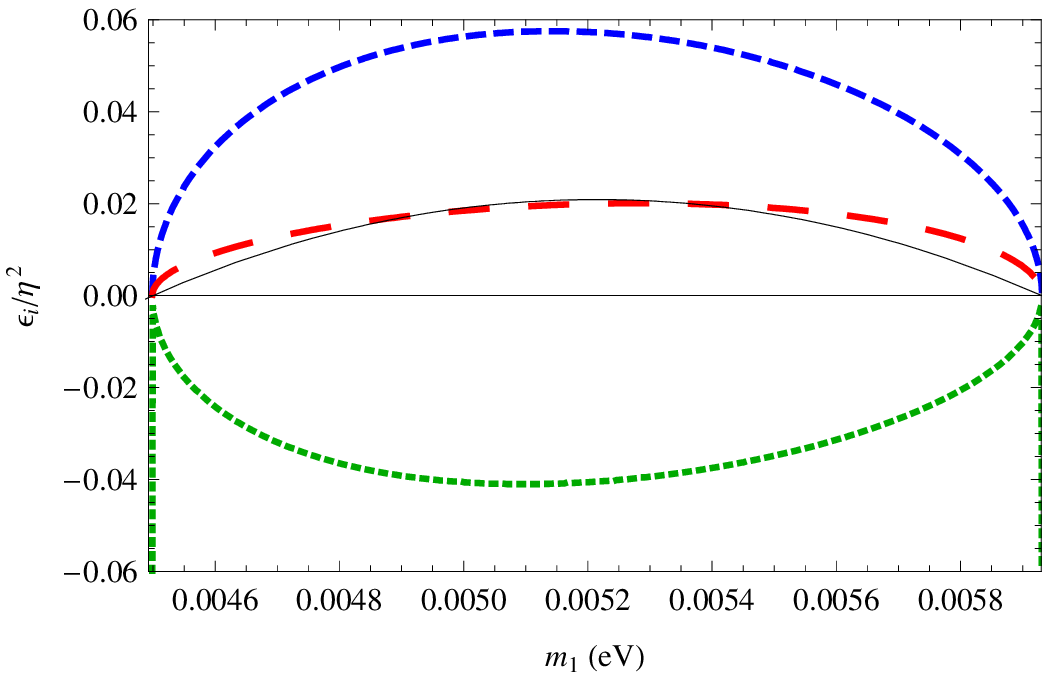,height=4.2cm,width=5.2cm,angle=0}
\hspace{-1mm}
\psfig{file=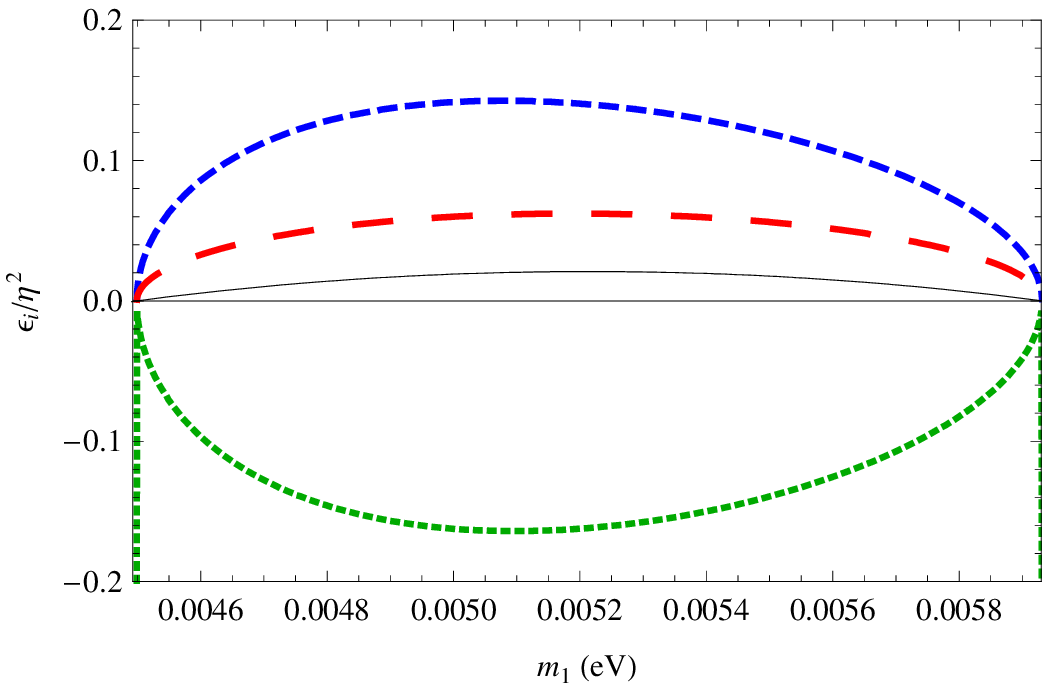,height=4.2cm,width=5.2cm,angle=0}
\\
\psfig{file=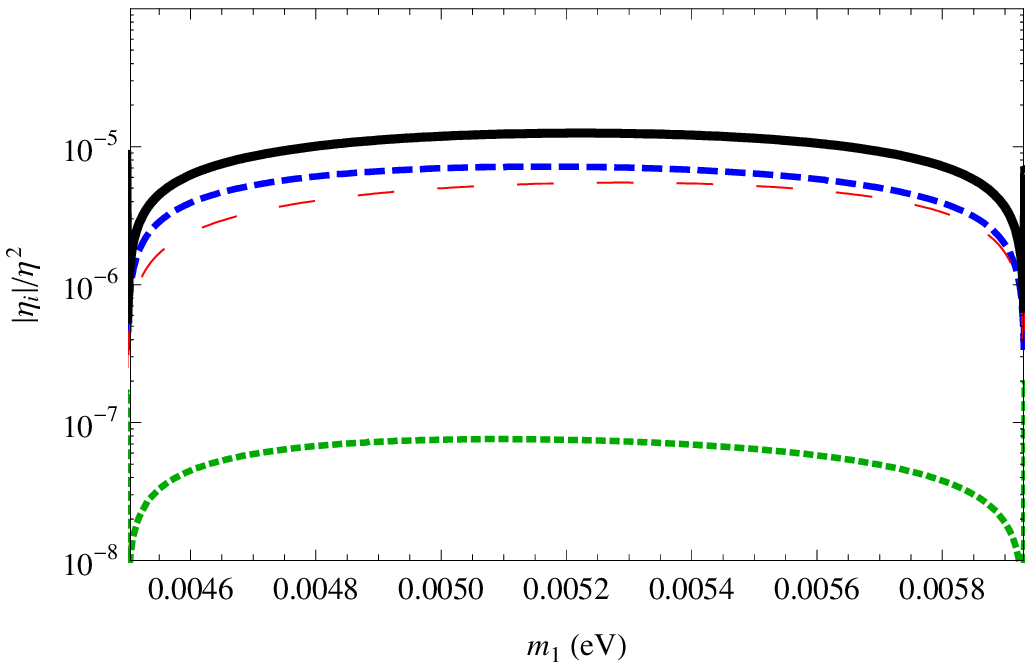,height=4.2cm,width=5.2cm,angle=0}
\hspace{-1mm}
\psfig{file=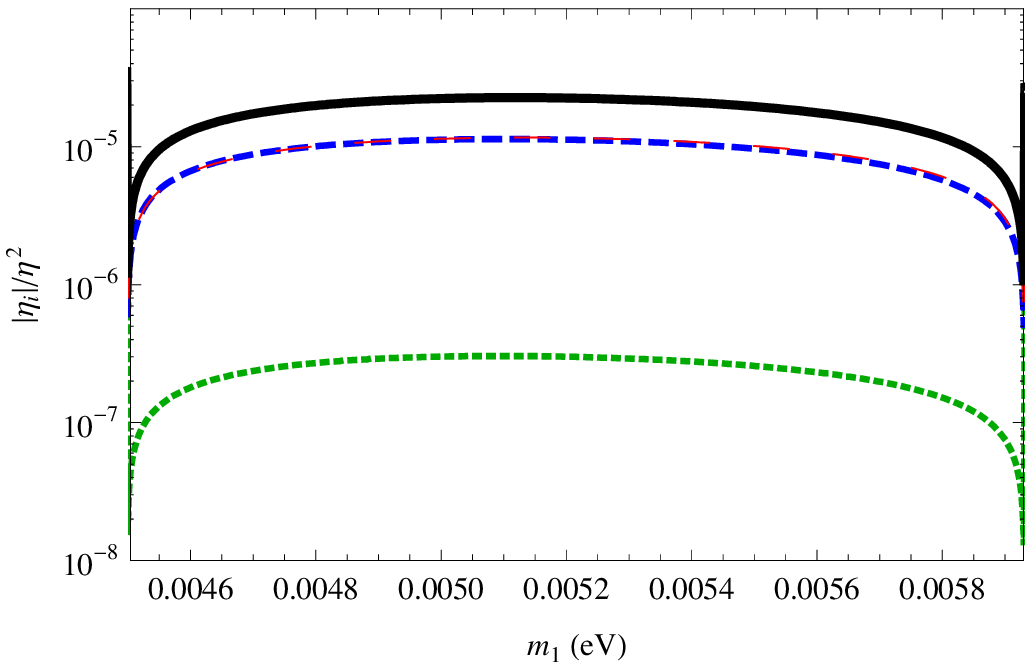,height=4.2cm,width=5.2cm,angle=0}
\hspace{-1mm}
\psfig{file=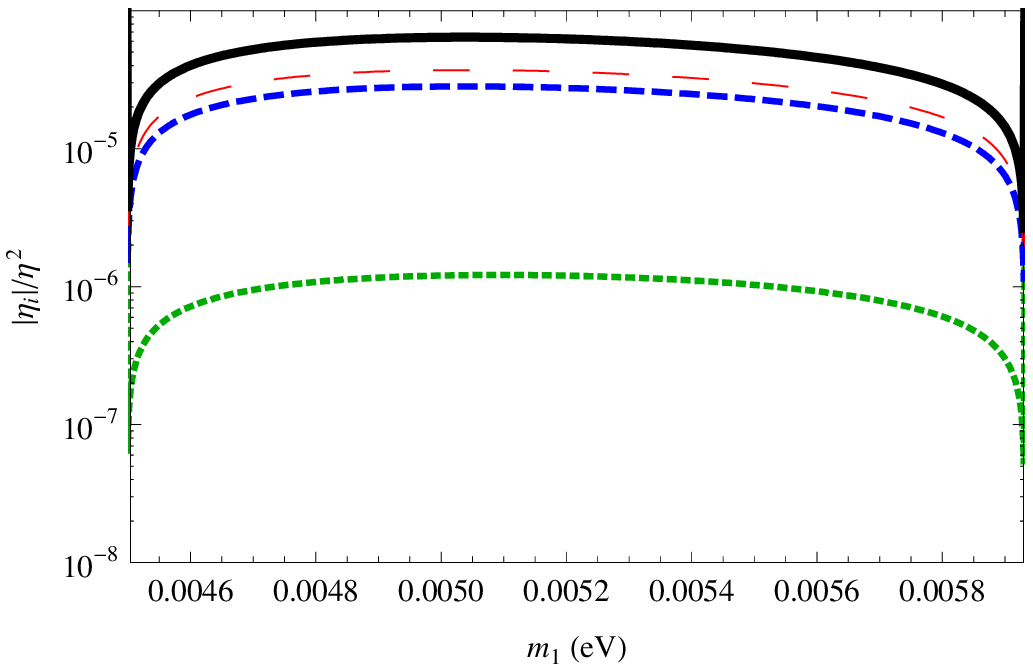,height=4.2cm,width=5.2cm,angle=0}
\\
\psfig{file=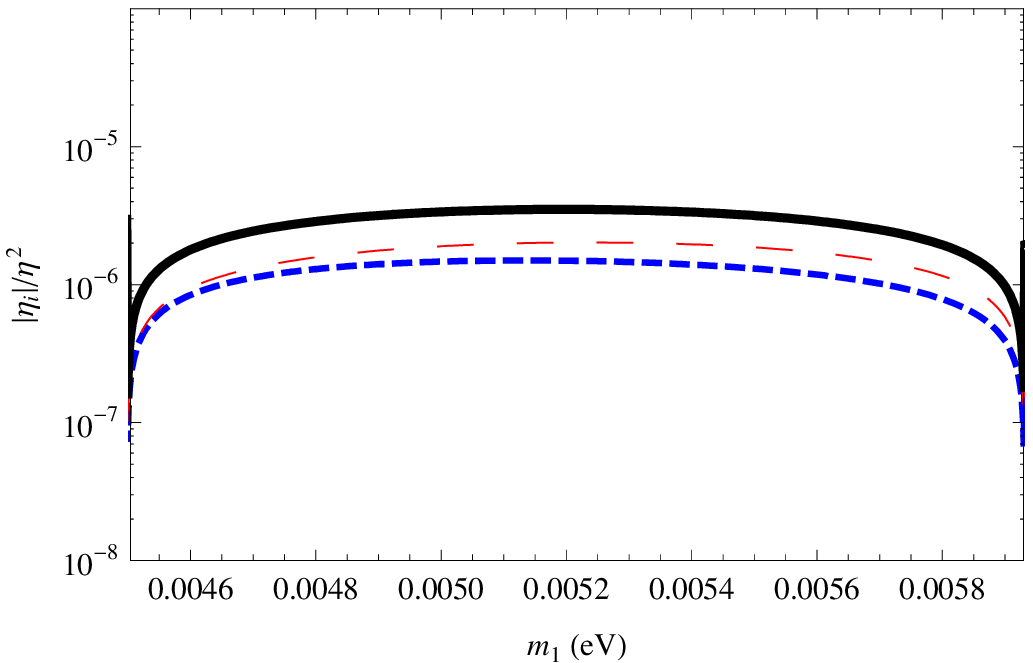,height=4.2cm,width=5.2cm,angle=0}
\hspace{-1mm}
\psfig{file=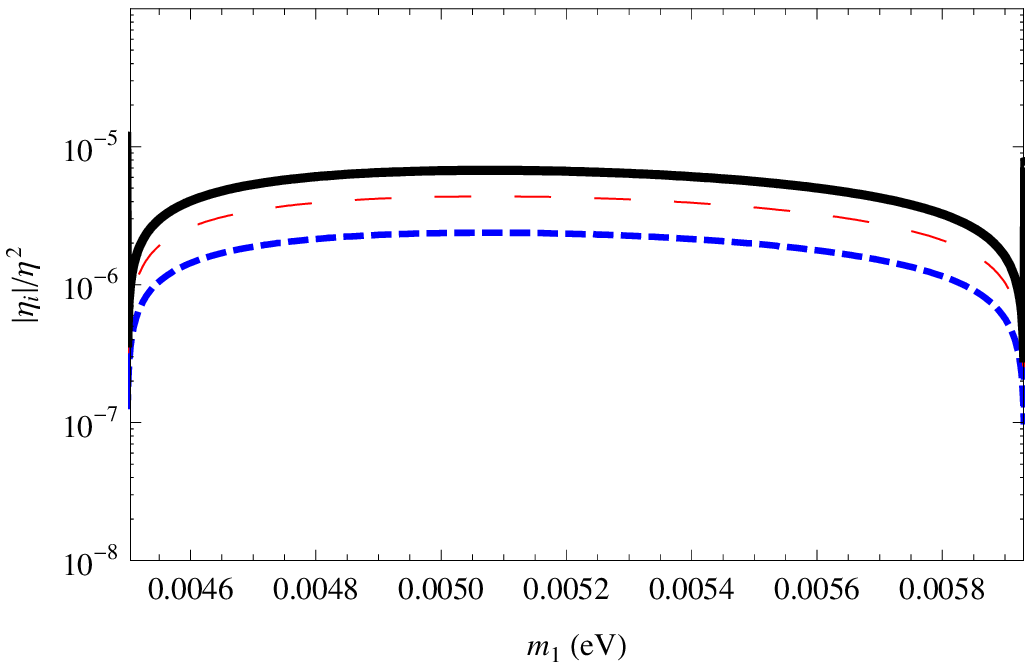,height=4.2cm,width=5.2cm,angle=0}
\hspace{-1mm}
\psfig{file=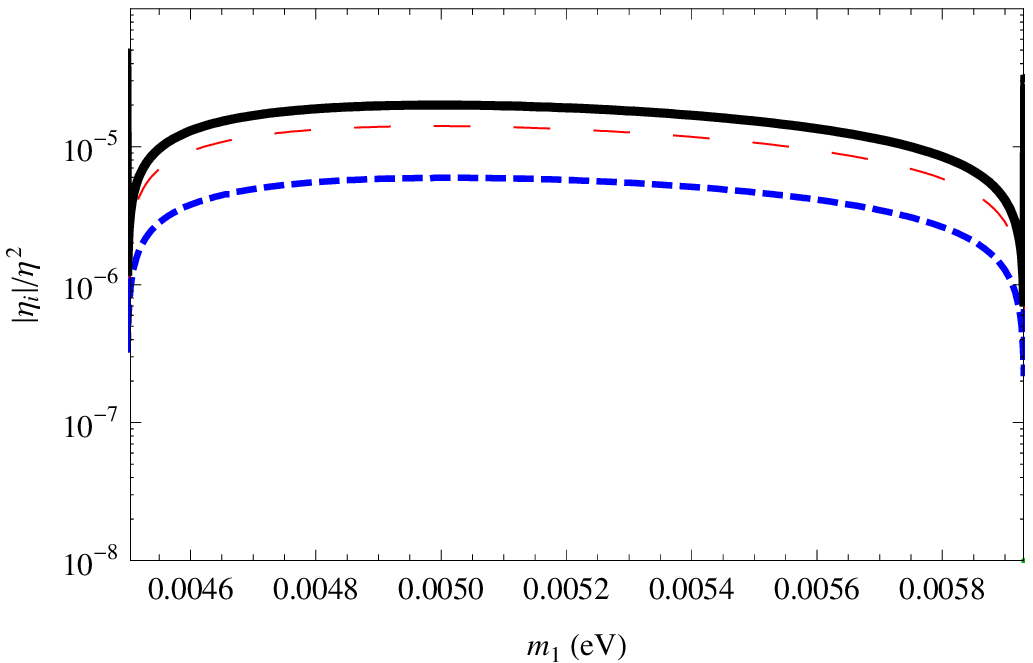,height=4.2cm,width=5.2cm,angle=0}
\\
\psfig{file=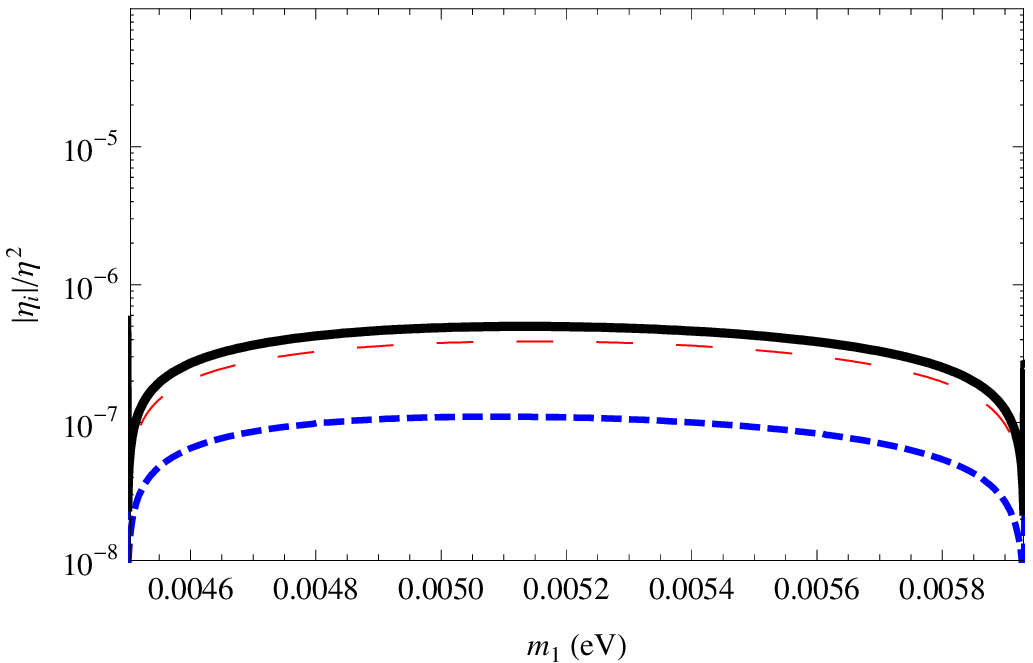,height=4.2cm,width=5.2cm,angle=0}
\hspace{-1mm}
\psfig{file=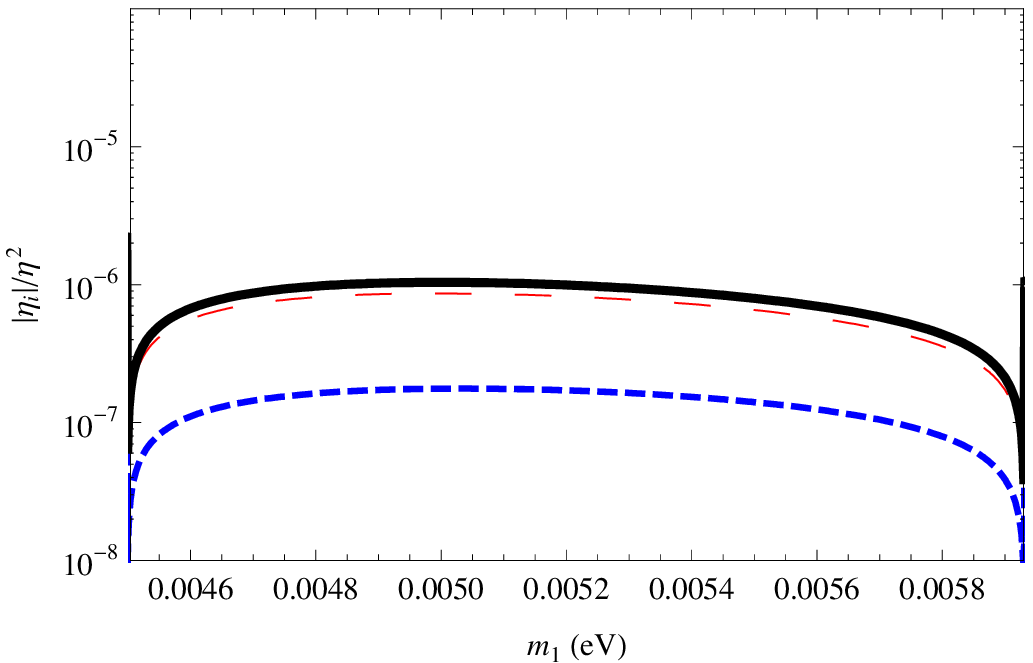,height=4.2cm,width=5.2cm,angle=0}
\hspace{-1mm}
\psfig{file=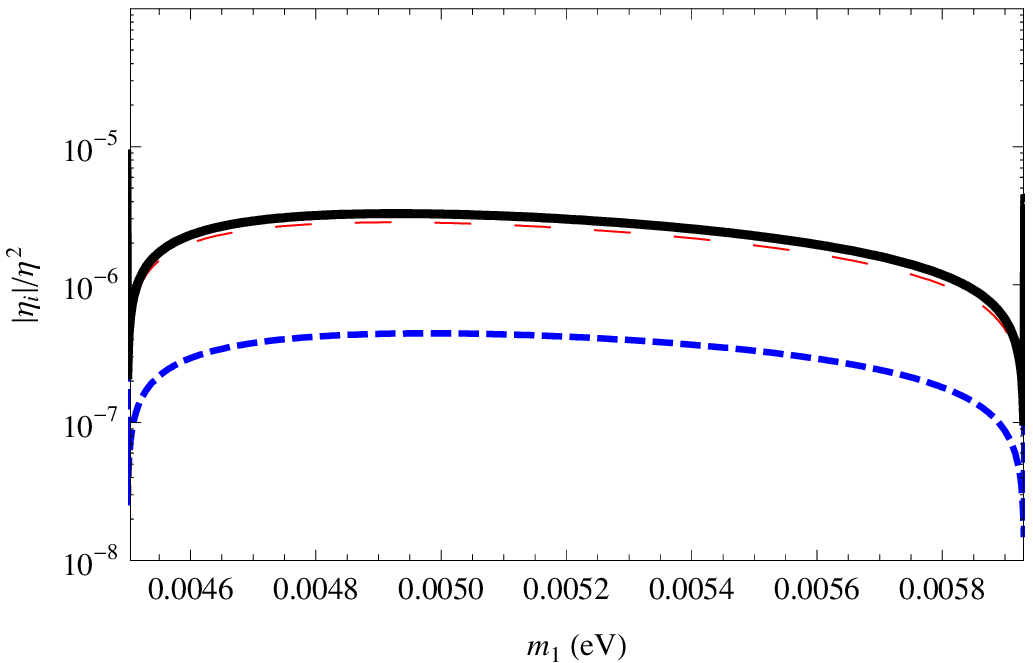,height=4.2cm,width=5.2cm,angle=0}
\caption{NO. In the 3 upper panels we plot the CP asymmetries $\ve_i$
  divided by $\eta^2$ as a function of $m_1$ for $\sin\varphi>0$ (note
  the different vertical scale in the last panel).  The dashed line
  corresponds to $i=1$, the short-dashed line to $i=2$ and the dotted
  line to $i=3$. The solid line is $1-\cos\varphi$
  (cf. Eq.~(\ref{phase})).  In the 9 lower panels we have plotted the
  three $|\eta_i|/\eta^2$ (cf. Eq.~(\ref{etaB2})), with the same line
  convention as for the $C\!P$ asymmetries for $i=1,2,3$ respectively,
  and $|\eta_B|/\eta^2$ (solid line).  All the left panels correspond
  to $\rho=0.5$, all the central panels to $\rho=1$ and all the right
  panels to $\rho=2$. The three rows of nine lower panels, showing the
  asymmetries, correspond to $y_\b=1$, $y_\b=2$ and $y_\b=3$ starting
  from above.}
\label{epsiNH}
\end{center}
\end{figure}

In the upper panels of Fig.~\ref{epsiNH} we show the dependence on
$m_1=m_l$ of the three CP asymmetries $\ve_i$ divided by the square of
symmetry breaking parameter $\eta^2$ for positive values of
$\sin\varphi$. For negative values they are simply all
opposite. Therefore, by switching the sign of $\sin\varphi$ the sign
of the final baryon asymmetry changes as well (i.e., the model does not
predict the sign of the baryon asymmetry).  It turns out that
the correct sign of the baryon asymmetry is obtained for
$\sin\varphi<0$.  In the lower panels of Fig.~\ref{epsiNH} we show the
absolute values of $\eta_{i}/\eta^2$ ($i=1,2,3$), that is the relevant
quantity, together with the total value $|\eta_B|=|\sum_i \eta_{i}|$.
We show the plots for three different values of $\rho=0.5, 1, 2$ from left to right and
for three different values of $y_\b=1,2,3$ from above to below. One can see how
for increasing $\rho$ typically the finally asymmetry increases.

On the other hand, increasing $y_\b$, the three RH neutrino masses increase
up to a critical value above which there is an exponential suppression from
$\Delta L=2$ processes. This critical value is about $y_\b\sim 2$ and
for $y_\b=3$ the value of $\eta_B/\eta^2$
reproduces the observed asymmetry for the highest possible values of the $\eta$-range
$5\times 10^{-3}\div 5\times 10^{-2}$.

It is also interesting to notice that the contribution from the
lightest RH neutrinos (depicted in Fig.~\ref{epsiNH} with the green
dotted line) is sub-dominant, despite the fact that the lightest RH
neutrino $C\!P$ asymmetry $|\ve_3|$ is typically the highest or anyway
comparable to $|\ve_2|$ (depending on the value of $m_1$). The reason
is that $K_3\simeq m_{\rm atm}/m_{\star}\gg K_{1,2}$ and therefore the
wash-out is much stronger compared to the two heavier RH neutrinos.

\begin{figure}
\begin{center}
\psfig{file=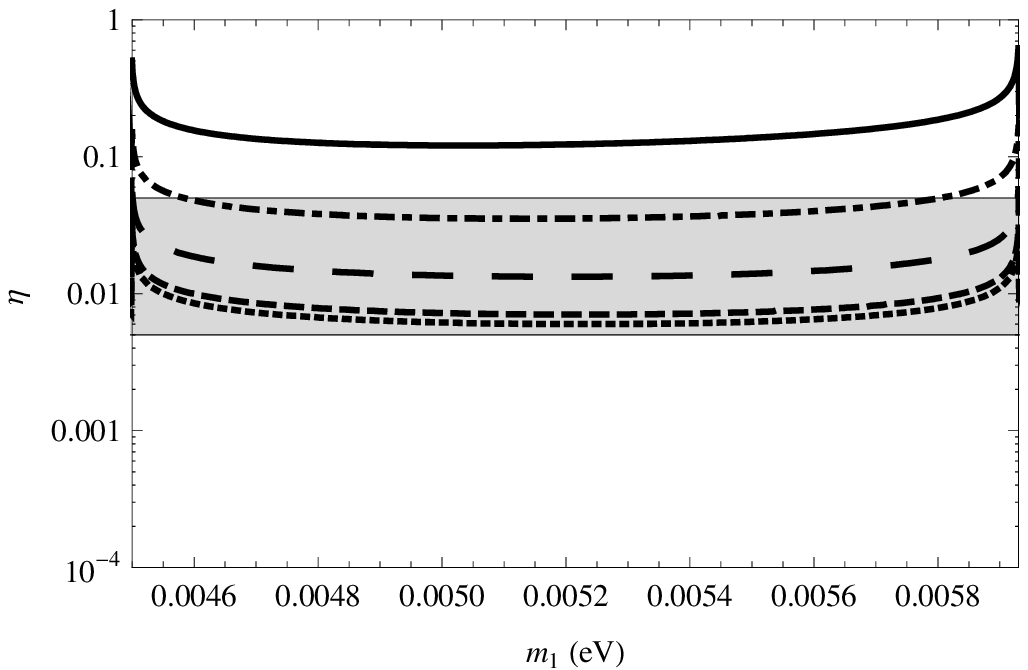,height=4.2cm,width=5.2cm,angle=0}
\hspace{-1mm}
\psfig{file=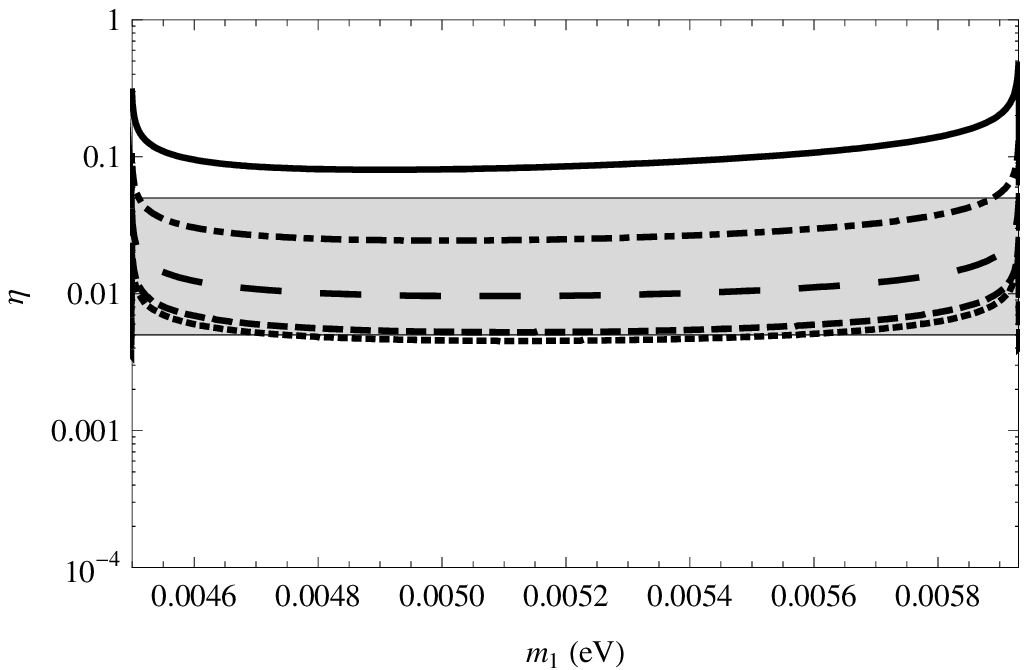,height=4.2cm,width=5.2cm,angle=0}
\hspace{-1mm}
\psfig{file=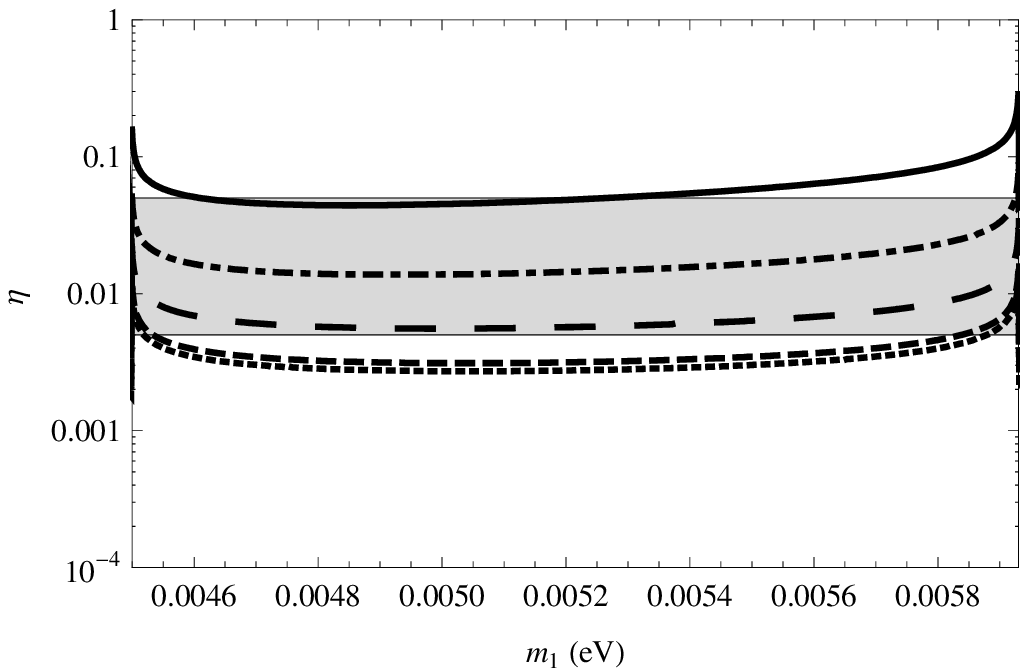,height=4.2cm,width=5.2cm,angle=0}
\caption{NO.  Plots of the values of $\eta$ that reproduce the
  observed $\eta_B$, as a function of $m_1$ for $\rho=0.5$ (left
  panel), $\rho=1$ (central panel), $\rho=2$ (right panel). The
  different curves correspond to $y_{\beta}=0.5$ (dotted), $y_\b=1$
  (short-dashed), $y_\b=2$ (long-dashed), $y_\b=3$ (dot-dashed) and
  $y_\b=4$ (solid). The gray band is the indicative optimal range of
  values of $\eta=5\times 10^{-3}\div 5\times 10^{-2}$.}
\label{etam1NH}
\end{center}
\end{figure}

In Fig.~\ref{etam1NH} we have plotted, as a function of $m_1$ and for different
choices of $\rho$ and $y_\b$, the value of $\eta$  such that $\eta_B=\eta_B^{CMB}$.
One can see that this value always falls in the optimal
range $\eta=5\times 10^{-3}\div 5\times 10^{-2}$ (the grey band).
It seems therefore that, despite the many constraints on the model
parameters and in particular the fact that there is only one
independent complex phase and that the single dimensional parameter
$m_1$ is practically fixed, the model reproduces the observed baryon
asymmetry for natural values of the parameters in quite a satisfactory
way.

\subsubsection{Inverted Ordering}

\begin{figure}
\begin{center}
\psfig{file=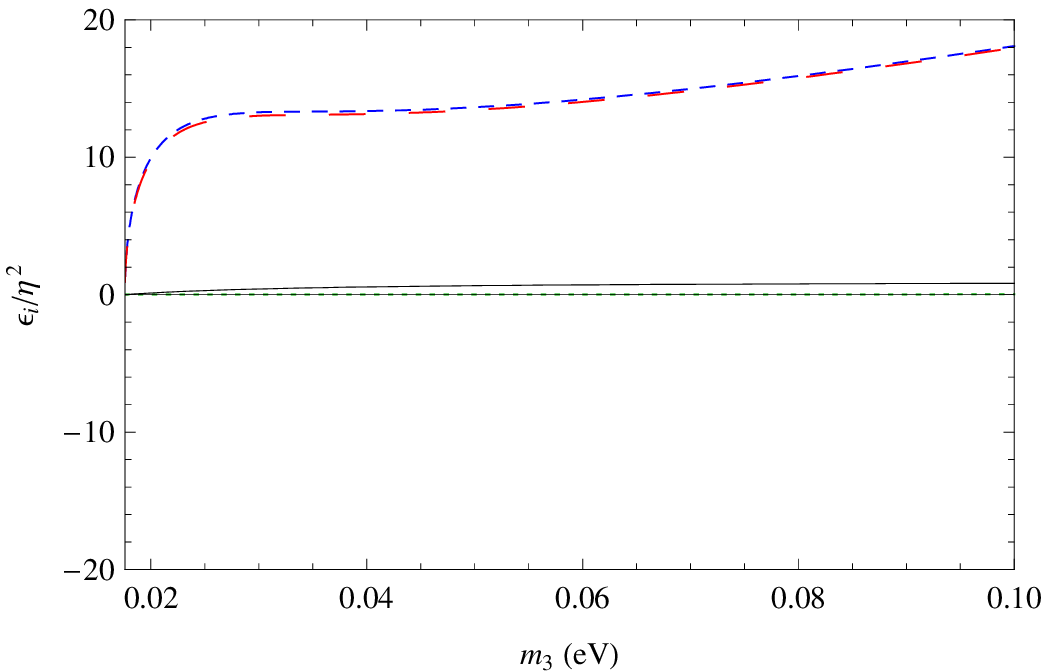,height=4.2cm,width=5.2cm,angle=0}
\hspace{-1mm}
\psfig{file=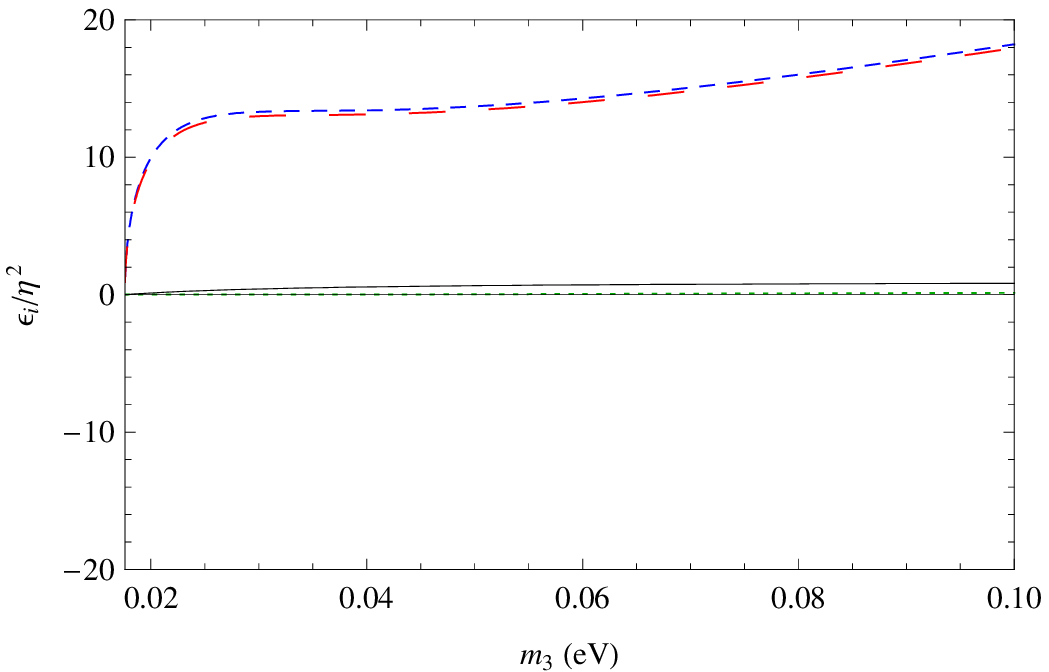,height=4.2cm,width=5.2cm,angle=0}
\hspace{-1mm}
\psfig{file=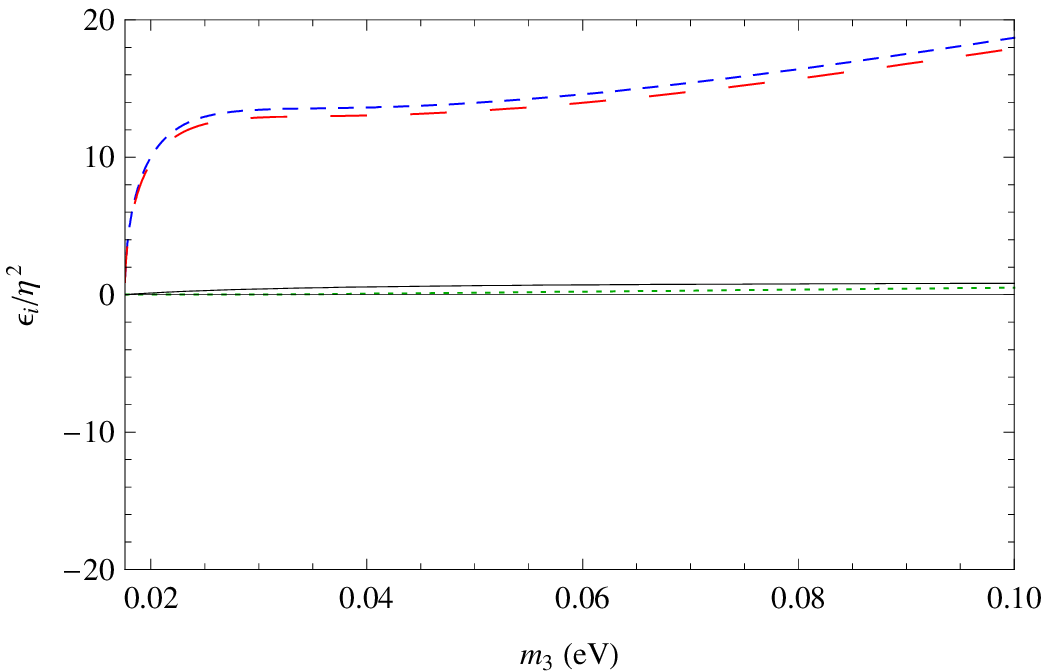,height=4.2cm,width=5.2cm,angle=0}
\\
\psfig{file=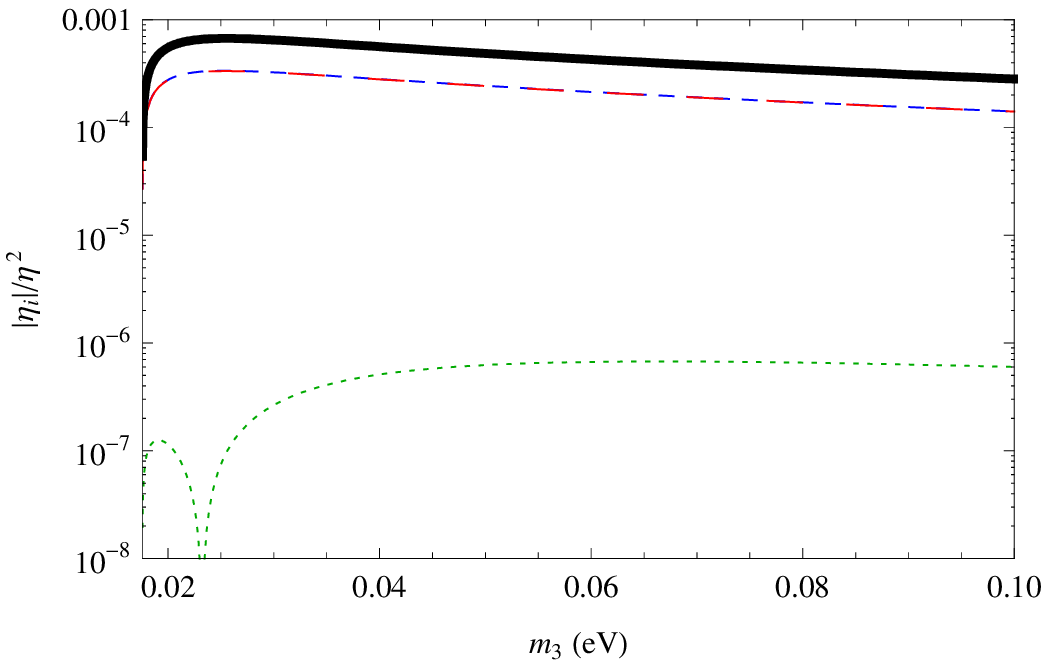,height=4.2cm,width=5.2cm,angle=0}
\hspace{-1mm}
\psfig{file=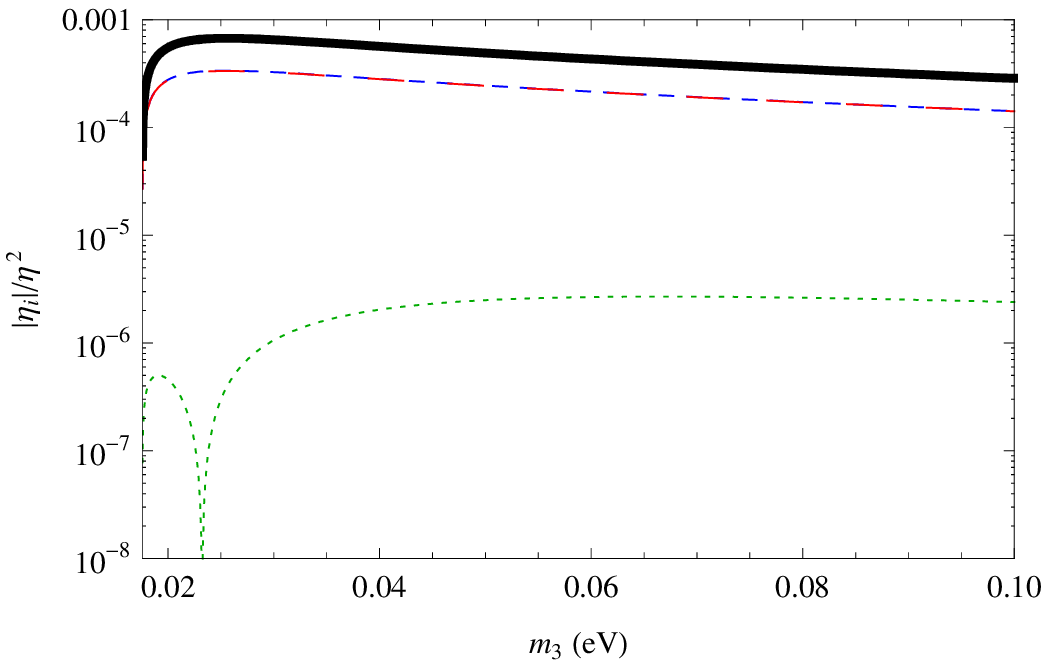,height=4.2cm,width=5.2cm,angle=0}
\hspace{-1mm}
\psfig{file=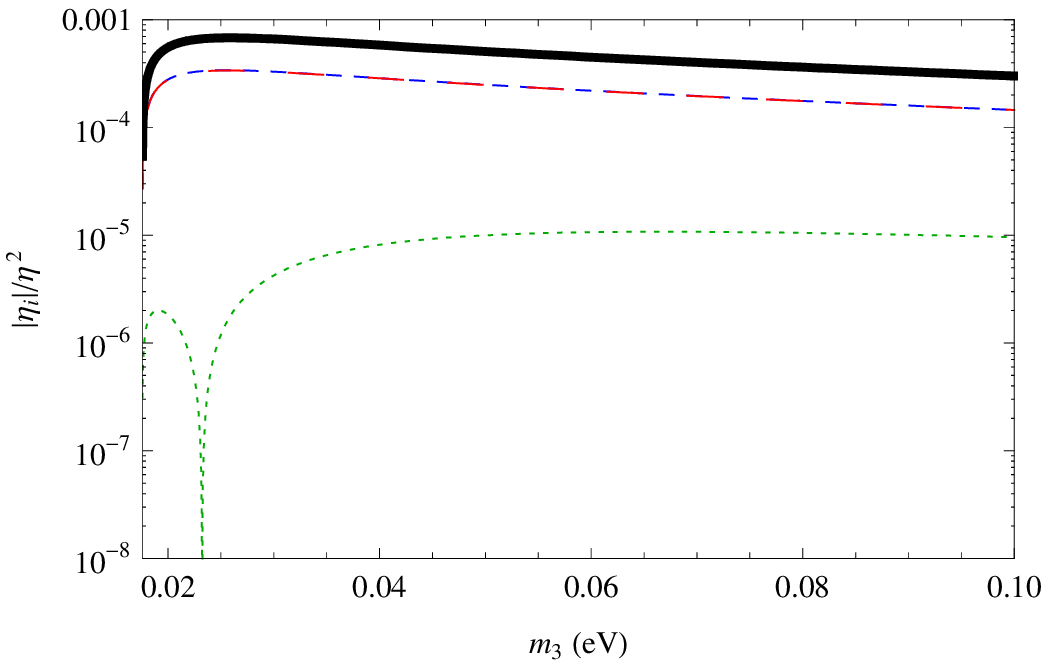,height=4.2cm,width=5.2cm,angle=0}
\\
\psfig{file=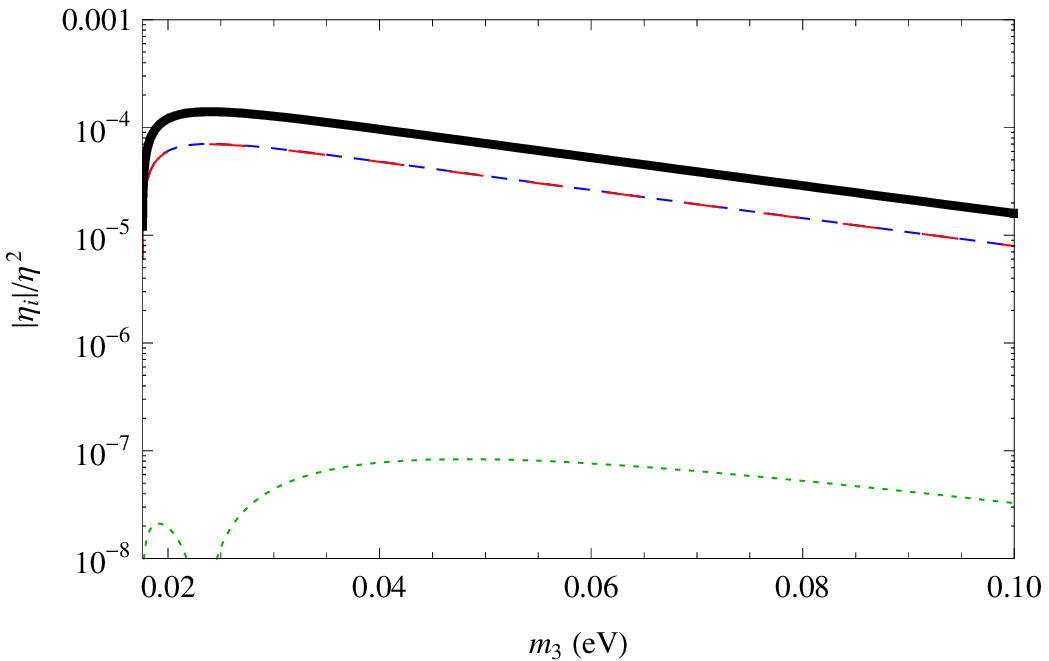,height=4.2cm,width=5.2cm,angle=0}
\hspace{-1mm}
\psfig{file=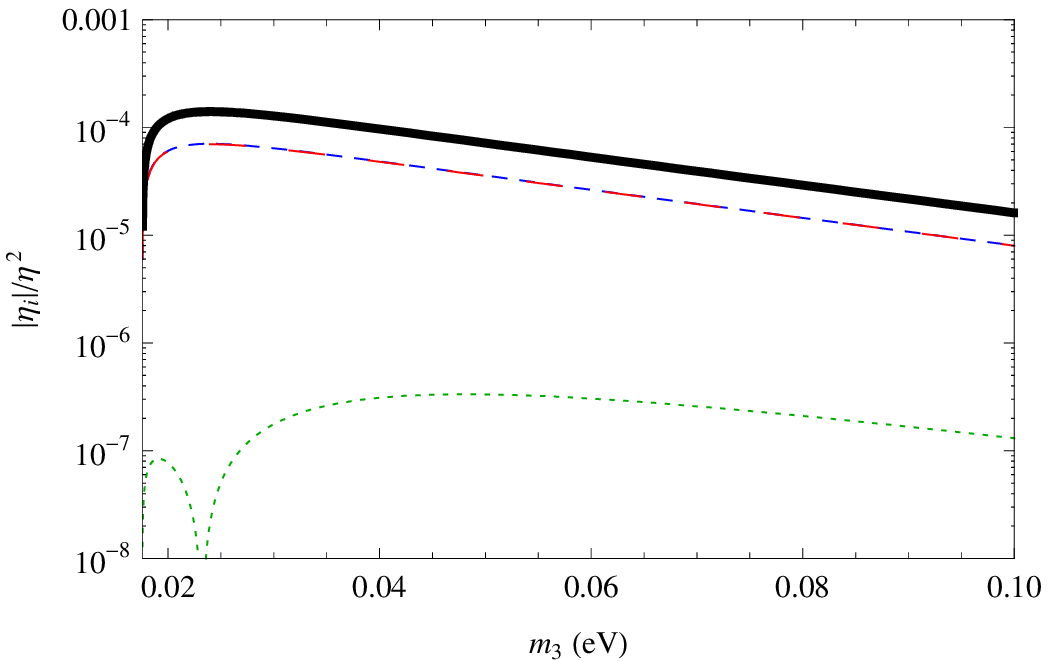,height=4.2cm,width=5.2cm,angle=0}
\hspace{-1mm}
\psfig{file=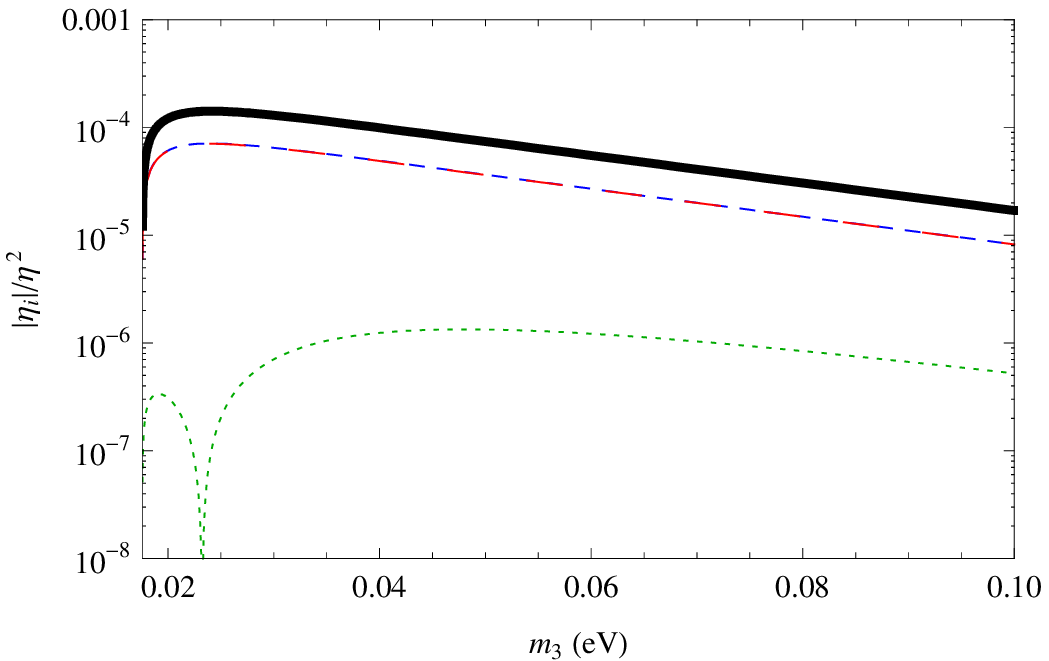,height=4.2cm,width=5.2cm,angle=0}
\\
\psfig{file=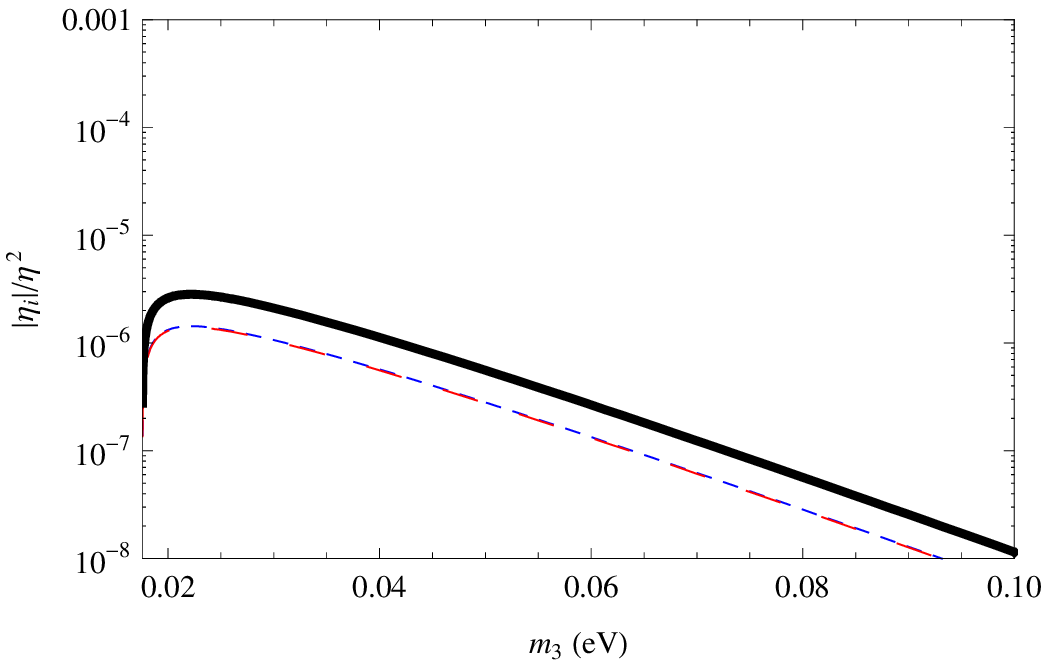,height=4.2cm,width=5.2cm,angle=0}
\hspace{-1mm}
\psfig{file=etaiIH171.eps,height=4.2cm,width=5.2cm,angle=0}
\hspace{-1mm}
\psfig{file=etaiIH171.eps,height=4.2cm,width=5.2cm,angle=0}
\caption{IO. In the 3 upper panels we plot the three $\ve_i/\eta^2$
as a function of $m_1$ and for $\sin\varphi>0$.
The dashed line corresponds to $i=1$, the short-dashed line to $i=2$ and the dotted
line to $i=3$. The solid line is $1+\cos\varphi$ (cf. Eq.~(\ref{phase})).
In the 9 lower panels we have plotted the three $|\eta_i|/\eta^2$
(cf. Eq.~(\ref{etaB2})) with the same line correspondence as for the
$C\!P$ asymmetries and $|\eta_B|/\eta^2$ (solid line).
All the left panels correspond to $\rho=0.5$, all the central panels to
$\rho=1$ and all the right panels to $\rho=2$. The three rows of nine lower panels showing the
asymmetries correspond to $y_\b=0.5$, $y_\b=1$ and $y_\b=1.7$ starting from above.}
\label{epsiIH}
\end{center}
\end{figure}

In the upper panels of Fig.~\ref{epsiIH}
we show the dependence on $m_3=m_l$ of the three  $\ve_i/\eta^2$
for positive values of $\sin\varphi$. Again, by switching the sign of $\sin\varphi$,
one can always change the final sign of the final asymmetry and again the correct
(positive) sign is obtained for $\sin\varphi< 0$.
We show again three different values of $\rho=0.5, 1, 2$ from left to right.
However, one can see that now there is only a tiny dependence on $\rho$.
The reason can be easily understood inspecting the three Eq.'s (\ref{asimmetria_CP}).
The strong degeneracy between the two lower RH neutrino masses, $M_1$ and $M_2$,
implies $|\xi(x_2/x_1)|, |\xi(x_1/x_2)|\sim 10^3$ and therefore $|f_{12}|, |f_{21}| \gg 1$.
In this way the term depending on $\rho$ gives a negligible contribution.

We again show examples for three different values of $y_\beta$. This
time we choose $y_\beta=0.5,1,1.7$ from above to below. One can see
how for $y_\beta=0.5$ one has $\eta_B/\eta^2 \sim 10^{-4}\div 10^{-3}$
and therefore too small values of $\eta\lesssim 10^{-3}$ are required
to explain the observed asymmetry. On the other hand for $y_\beta=1.7$
one has $\eta_B/\eta^2 \sim 10^{-8}\div 10^{-6}$ and this time too
large values of $\eta\lesssim 5\times 10^{-2}\div 5\times 10^{-1}$ are
needed.  Therefore, in these examples, the observed asymmetry is
reproduced for reasonable values of $\eta\sim 0.01$ only for
$y_\beta=1$. For this value there is indeed a compensation between
very large values of the $C\!P$ asymmetries, $|\ve_i|\sim 10^{-2}$ for
$\eta\sim 10^{-2}$, and an additional wash-out suppression $\sim
10^{-4}$ coming from $\Delta L=2$ processes so that the resulting
efficiency factor $\sim 10^{-6}$.

\begin{figure}
\begin{center}
\psfig{file=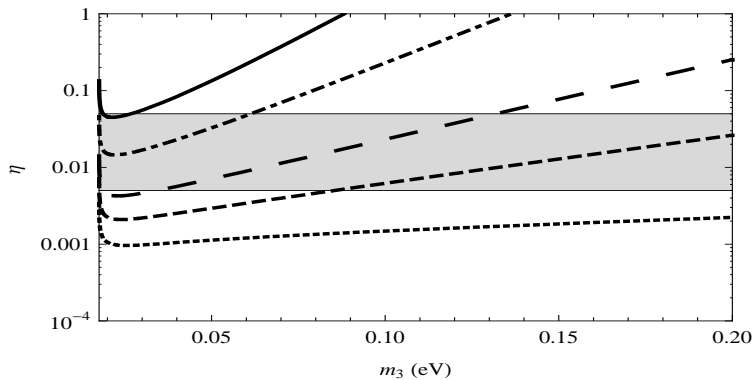,height=5cm,width=10cm,angle=0}
\caption{IO. Plot of the values of $\eta$ that reproduce the observed
  $\eta_B$, as a function of $m_1$ for $\rho=1$. The different curves
  correspond to $y_{\beta}=0.5$ (dotted), $y_\b=1$ (dashed),
  $y_\b=1.3$ (long-dashed), $y_\b=1.7$ (dot-dashed) and $y_\b=2$
  (solid). The gray band is the indicative optimal range of values of
  $\eta=5\times 10^{-3}\div 5\times 10^{-2}$.}
\label{etam1IH}
\end{center}
\end{figure}

We have again summarized the situation plotting in  Fig.~\ref{etam1IH},
the value of $\eta$ such that $|\eta_B|=\eta_B^{CMB}$ as a function of $m_l$.
We show  five examples for $y_\b=0.5, 1, 1.3, 1.7, 2$.
One can see that this time $\eta$ naturally falls in the optimal
range $\eta=5\times 10^{-3}\div 5\times 10^{-2}$ (the grey band) only for
for $y_\b=1\div 1.7$, therefore requiring some amount of tuning.
Moreover this occurs not for all values of $m_l$ and in particular
values $y_\b\simeq 1$ require $m_l\gtrsim 0.1\,{\rm eV}$.  Therefore,
an improvement of the upper bound on $m_l$ (cf. (\ref{upperbound}))
will enforce higher values of $y_\b$ corresponding to higher values of
the RH neutrino masses that in turn imply higher values of the initial
temperature of the radiation dominated regime that, in an inflationary
language, corresponds to the reheating temperature of the Universe.

\subsection{Reheating temperature constraints}

The model predicts a RH neutrino spectrum with a mild hierarchy
and an overall scale of values of the RH neutrino masses that is quite
large. For NO one has $M_{3} \sim 10^{15}\,{\rm GeV}\,y^2_\b$, where $M_3$
is the mass of the lightest RH neutrino. For IO the scale can be
about five times lower if one considers quasi-degenerate light neutrinos
with masses $m_i\gtrsim 0.1\,{\rm eV}$, close to the current cosmological
upper bound (cf. (\ref{upperbound})).

Since RH neutrinos are produced by thermal processes, this in turn
implies a lower bound on the reheating temperature given approximately
by \cite{pedestrians} $T_{\rm reh}\gtrsim M_{\rm 3}/[z_B(K_{\rm
  3})-2]\simeq 10^{14}\,y^2_\b\,{\rm GeV}$ in the case of NO and about
five times lower for IO.  Are such high values of the $T_{\rm reh}$
possible ?  Since our model is supersymmetric, the well known upper
bound $T_{\rm reh}\lesssim 10^{6\div 10}\,{\rm GeV}$ from the
avoidance of the gravitino problem potentially applies
\cite{gravitino}. It is clear that such low values cannot be obtained
in the presented version of the model since, even taking the low value
$y_\b\sim 0.1$, one obtains $T_{\rm reh}\gtrsim 10^{12}\,{\rm GeV}$
for NO.

There are two possible kinds of ways out remaining within thermal
leptogenesis. The first one would be to
circumvent the gravitino problem and indeed
a few solutions have been proposed  \cite{buchmuwi}.
The second kind would be to modify the model in such a way that the
overall RH neutrino mass scale is lowered. It would not be difficult
to envisage different schemes to this extent.

On the other hand, in the case of IO  the second strategy  cannot
be invoked for the simple reason that large values of the RH neutrino masses
$M_l\gtrsim 10^{14}\,{\rm GeV}$ are necessary in order to get a strong
additional wash-out suppression from $\D L=2$ processes to compensate
the very large values of the $C\!P$ asymmetries. In other words large reheating
temperatures $T_{\rm reh}\gtrsim 5\times 10^{13}\,{\rm GeV}$ cannot be avoided
in the case of IO.

\section{Conclusions}

A typical outcome of extensions of the standard model that attempt to explain the features
of the lepton mass spectrum on the basis of flavor symmetries is that
small quantities such as the charged lepton mass ratios, $\Delta m^2_{sol}/\Delta m^2_{atm}$, $\theta_{23}-\pi/4$ and $\theta_{13}$
are proportional to some power of small symmetry breaking parameters $\eta$.
By keeping only the leading order power of $\eta$, in some cases the number of independent parameters
becomes small and these models can be rather predictive, with characteristic  relations among
the observable quantities. Well known examples are relations between the neutrino oscillation parameters
and the branching ratios of lepton flavor violating processes, such as $\mu \to e \gamma$, $\tau\to \mu\gamma$ and $\tau\to e \gamma$.
It would be quite interesting to include in this kinships also
the baryon asymmetry, which, in the context of leptogenesis, is naturally related to lepton masses and mixing angles.

With such scope in our mind, we have discussed the constraints on the leptogenesis $C\!P$ asymmetries in models possessing a flavor symmetry.
We have derived general conditions for the vanishing of the $C\!P$
asymmetries in the limit of exact flavor symmetry.
We have shown that, if the three RH neutrinos belong to
an irreducible representation of the flavor symmetry group, then the total $C\!P$ asymmetries are zero
in the limit of exact flavor symmetry. More precisely, for non-degenerate RH neutrino masses, the total $C\!P$ asymmetries
$\epsilon_i$ are of order $\eta^2$ and the flavored ones $\epsilon_{i\to \alpha}$ are of order
$\eta$.
If the RH neutrinos are not in an irreducible representation of the flavor group,
we have derived a necessary and sufficient condition for the vanishing of the total $C\!P$ asymmetry
in the symmetric limit and we have discussed it in several particular cases. For instance,
if the action of the symmetry on RH neutrinos is Abelian, then in most cases
the $C\!P$ asymmetries are of order $\eta^0$ and we should invoke additional
washout suppression to reproduce the observed baryon asymmetry.

One interesting example of vanishing leading-order $C\!P$ asymmetries is that of a model symmetric under
$A_4\times Z_3\times U(1)_{FN}$, built to reproduce tri-bimaximal
lepton mixing. In this model RH neutrinos are in a triplet of $A_4$ and
$\epsilon_i=O(\eta^2)$. The model is rather constrained. Once the parameters are fixed
to match $\Delta m^2_{sol}$ and $\Delta m^2_{atm}$, there is only one relevant phase $\varphi$, which can be
thought of as a function of the lightest neutrino mass $m_l$. The RH neutrino spectrum
depends only on an additional, $O(1)$, parameter $y_\beta$.
Both normal and inverted neutrino mass ordering can be reproduced.
For normal ordering $m_l$ is essentially fixed in a small range around $0.005$ eV
and the phase should be very small. For inverted hierarchy there is a lower bound on $m_l$
of approximately $0.017$ eV and there is much more freedom for the phase. Given this rather constrained framework it is not guaranteed
that a successful leptogenesis can take place at all and we computed the
washout effects and the resulting baryon asymmetry. The dynamics of the model is quite
interesting since the RH neutrinos have similar masses and they all participate to generate
the baryon asymmetry. At the same time, to a good approximation, the asymmetry produced in the
decay of one RH neutrino is not washed-out by the other heavy neutrino inverse processes,
since the interactions of the three RH neutrinos with the light leptons are almost orthogonal to each other.
For normal hierarchy we find that  the observed baryon asymmetry is reproduced
for values of the symmetry breaking parameter $\eta$ in the range $0.005\div 0.05$,
which nearly coincides with the natural expected range in this model. This prediction is rather stable
with respect to variation of $y_\beta$. For inverted hierarchy
we find solutions in the parameter space, but they are less stable. If $0.005<\eta<0.05$,
there is only a rather limited range of allowed values for $y_\beta$.
    On the one hand, for small values of $y_\beta$ the baryon asymmetry is
typically enhanced, compared to the normal ordering case and we cannot go below $y_\beta\simeq 1$,
with a reheating temperature not lower than $5\times 10^{13}$ GeV.
   On the other hand, as soon as $y_\beta$ exceeds 2
the suppression from the washout becomes huge and the baryon asymmetry goes rapidly to zero.

It is interesting that though the IO case has a much wider parameter freedom
compared to the NO case, it certainly appears less attractive from a cosmological
point of view. Future improvements on the measurements of the
absolute neutrino mass scale will allow to test our results.

\vspace{3mm}

\textbf{Acknowledgments} We warmly acknowledge Emiliano Molinaro and
Serguey Petcov for pointing out a mistake in eq.~(\ref{asimmetria_CP})
in the first version of the paper.  PDB was supported by the Helmholtz
Association of National Research Centres, under project VH-NG-006.
The work of EN is supported in part by Colciencias under contract
1115-333-18739.

\end{document}